\documentclass[%
 aip,
 amsmath,amssymb,
]{revtex4-1}
\usepackage{verbatim}
\usepackage{graphicx}
\usepackage{dcolumn}
\usepackage{bm}

\usepackage[utf8]{inputenc}
\usepackage[T1]{fontenc}
\usepackage{mathptmx}
\usepackage{etoolbox}
\usepackage{physics}
\newcommand*\diff{\mathop{}\!\mathrm{d}}
\newcommand*\Diff[1]{\mathop{}\!\mathrm{d^#1}}
\makeatletter
\def\@email#1#2{%
 \endgroup
 \patchcmd{\titleblock@produce}
  {\frontmatter@RRAPformat}
  {\frontmatter@RRAPformat{\produce@RRAP{*#1\href{mailto:#2}{#2}}}\frontmatter@RRAPformat}
  {}{}
}%
\makeatother
\begin{document}

\preprint{AIP/123-QED}

\title[]{Development and application of a hybrid MHD-kinetic model in JOREK}
\author{T.J. Bogaarts} 
\affiliation{Eindhoven University of Technology, PO Box 513, 5600 MB, Eindhoven, Netherlands}%
\affiliation{Max-Planck-Institute für Plasmaphysik, D-85748, Garching, Germany}
\author{M. Hoelzl}%
\affiliation{Max-Planck-Institute für Plasmaphysik, D-85748, Garching, Germany}
 \email{mhoelzl@ipp.mpg.de}

\author{G.T.A. Huijsmans}
\affiliation{Eindhoven University of Technology, PO Box 513, 5600 MB, Eindhoven, Netherlands}
\affiliation{CEA, IRFM, F-12108, Saint-Paul-lez-Durance, France}
\author{X. Wang}
\affiliation{Max-Planck-Institute für Plasmaphysik, D-85748, Garching, Germany}
\author{JOREK team}
\affiliation{See the author list of M. Hoelzl et al, Nucl. Fusion, vol. 61, p. 065001, 2021}

\date{\today}

\begin{abstract}
Energetic particle (EP) driven instabilities will be of strongly increased relevance in future burning plasmas as the EP pressure will be very large compared to the thermal plasma. Understanding the interaction of EPs and bulk plasma is crucial for developing next-generation fusion devices. In this work, the JOREK MHD code and its full-f kinetic particle-in-cell module is extended by an anisotropic pressure coupling model to allow for the simulation of EP instabilities at high EP pressures using realistic plasma and EP parameters. Furthermore, a diagnostic is implemented to allow for the visualisation of phase-space resonances. The resulting code is first benchmarked linearly for the ITPA-TAE as well as the experiment based AUG-NLED cases, obtaining good agreement to other codes. Then, it is applied to a high energetic particle pressure discharge in the ASDEX Upgrade tokamak using a realistic non-Maxwellian distribution of EPs, reproducing aspects of the experimentally observed instabilities. Non-linear applications are possible based on the implentation, but will require dedicated verification and validation left for future work. 
\end{abstract}

\maketitle

\section{\label{sec:intro} Introduction}
In nuclear fusion reactors, energetic particles (EPs), with characteristic energies much larger than the thermal energy of the bulk plasma, can arise due to fusion reactions or external heating systems. Confining these energetic particles is crucial for sustaining a burning fusion reaction. These EPs can interact resonantly with magnetohydrodynamic (MHD) waves or instabilities of the bulk plasma, leading to outwards transport and possibly deconfinement of EPs. In present-day devices, EPs normally only provide a small fraction of the total pressure, while in future burning reactors, the EP pressure is high compared to the bulk plasma. Thus, for predicting the performance and optimizing the design of future fusion devices, understanding the interaction of EPs with the bulk plasma in a regime with high EP pressure (but not necessarily high EP density) is a key area of research. 

For simulating the interaction of EPs with a bulk plasma, a common technique is the hybrid MHD-kinetic model\cite{park1992three}, employed by for example MEGA\cite{todo1998linear}, (X)HMGC \cite{briguglio1995hybrid,wang2011extended}, XTOR-K\cite{leblondthesis},  M3D-K\cite{shen2014m3d}, M3D-C1-K\cite{liu2022hybrid}. Here, the bulk plasma is treated using an MHD model, while the EPs are treated kinetically. This approach saves computational time compared to a fully kinetic treatment such as implemented in ORB5\cite{lanti2020orb5}, but can still reproduce the relevant physics\cite{vlad2021linear}.

In this work, a hybrid MHD-kinetic extension of the non-linear extended MHD code JOREK\cite{hoelzl2021jorek} is introduced, capable of simulating EP driven instabilities in a high EP pressure discharge using realistic, experimental plasma parameters and EP distribution functions. A full-f formulation is employed for the EPs and an anisotropic pressure coupling to the MHD fluid is used. The JOREK code contains a broad range of different MHD models, can simulate up to the first wall, and has proven capabilities for challenging, highly non-linear MHD scenarios. It also has a versatile kinetic particle module, capable of simulating many different types of particles, varying from slow impurities to relativistic electrons. This module is currently being ported to graphical processing units (GPUs)\cite{huijsmansprivatecom}. These features make it an attractive option for the simulation of a wide variety of EP instabilities. Earlier work\cite{dvornova2021hybrid} used the JOREK kinetic extension for Toroidal Alfvén Eigenmodes (TAEs)\cite{cheng1985high} in the presence of EPs, but the model used therein is generalised to anisotropic EP distribution functions in the present work.

This hybrid extension of JOREK is benchmarked to other codes for TAEs and Energetic Particle Modes (EPMs)\cite{chen1994theory}, with good agreement regarding mode structures, frequencies, growth rates, and EP phase space resonances in the linear regime. The code is then applied to a high EP pressure discharge in the ASDEX-Upgrade (AUG) tokamak to validate the model and show its potential for simulating scenarios relevant to ITER and DEMO.

The paper is structured as follows. In section \ref{sec:JOREK} the JOREK code and the EP simulation model is introduced. In section \ref{sec:phase_space} the phase-space diagnostic in JOREK is explained. Linear benchmarks with other EP-simulation capable codes are shown in section \ref{sec:benchmarks}. Results from the application of the developed code to a high EP pressure discharge in the ASDEX-Upgrade (AUG) tokamak are shown in section \ref{sec:application} and a summary and outlook is finally given in section \ref{sec:conclusion}.

\section{\label{sec:JOREK} Energetic Particle model in JOREK}
The JOREK code\cite{hoelzl2021jorek} is a 3D non-linear extended MHD code, capable of simulating tokamak plasmas in realistic X-point geometry using a broad range of models. It uses implicit time-stepping, a finite element discretisation in the poloidal plane and a Fourier series expansion in the toroidal angle. Reduced and full MHD models with various extensions are avaible. A comprehensive overview of the models, numerics, and applications is available in Ref. \onlinecite{hoelzl2021jorek}.

A kinetic extension of JOREK was developed initially for the simulation of edge impurities during edge localised modes\cite{van2019nonlinear} but has been extended for other applications such as runaway electrons \cite{sarkimaki2022confinement}, edge physics, and ion temperature gradient turbulence studies \cite{huijsmansprivatecom}. It uses a particle-in-cell scheme to solve the kinetic Boltzmann equation for a particular species, which can be coupled to the MHD fluid if desired. The kinetic particle module is coupled to the MHD fluid by projecting moments of the distribution function on the finite element representation of this MHD fluid and using these projections in the timestepping of the MHD fluid. The specific terms coupled to the MHD fluid can depend on the application (e.g. collisionless EPs have different coupling terms than heavy impurities).

A 'full-f' Particle-In-Cell scheme is used for the solution of the distribution function of the EPs, such that the distribution function $f$ for $N_p$ marker particles is 
\begin{equation}\label{eq:fullf}
    f(\mathbf{r},\mathbf{v},t) = \sum_{i=1}^{N_p} w_i \delta[\mathbf{r}-\mathbf{r}_i]\delta[\mathbf{v}-\mathbf{v_i}],
\end{equation}
where $w_i$ indicates the number of physical particles the $i$th marker particle represents. The marker particles are pushed using one of the available pushers, ranging from full-orbit to relativistic gyro-centre. Collisions with the bulk fluid and the kinetic particles can be used, and ionisation, recombination, and radiation models are included. The particle pushing is parallelized on central processing units (CPUs) using hybrid OpenMP-MPI while for GPU accelaration OpenACC is used.  Marker particles can be initialised using analytical profiles for spatial and velocity space distribution functions, or from arbitrary, numerical distribution functions, e.g., realistic distribution produced by the NUBEAM code \cite{pankin2004tokamak}.

For EPs, collisions are neglected and a full-orbit Boris pusher\cite{delzanno2013particle} is used to retain all finite orbit width (FOW) and finite Larmor radius (FLR) effects. Guiding and gyro-center pushers are available and may be used in future applications. The coupling between EPs and the bulk MHD fluid is provided by the pressure coupling scheme\cite{park1992three} in the MHD momentum equation as 
\begin{equation}
    \rho \left(\pdv{\textbf{V}}{t}+\vb{V} \cdot \nabla \vb{V}\right)+\nabla p  -\vb{J}_t \times \vb{B} =  - \underline{\left(\nabla \cdot \vb*{\Pi}_h\right)_\perp}, 
\end{equation}
where 
\begin{equation*}
    \vb*{\Pi}_h = m_h \int \vb{v}\otimes \vb{v} f_h \Diff3\vb{v} 
\end{equation*}  

is the pressure tensor of the EPs calculated from the EP distribution function $f_h$. $\rho$ is the bulk fluid mass density, $\vb{V}$ is the bulk fluid velocity and $p$ the bulk fluid pressure. $\vb{J}_t$ is the total current carried by the bulk plasma and EPs, and $\vb{B}$ the magnetic field. In the latter equation, $m_h$ is the EP mass and $\vb{v}$ the velocity coordinate of the distribution function $f_h$. The symbol $\perp$ denotes the component perpendicular to  the magnetic field. The current coupling scheme has been implemented as well, see Ref. \onlinecite{dvornova2021hybrid}, but was too noisy to obtain useful results. In a previous simplified implementation\cite{dvornova2021hybrid}, only scalar pressure was considered. Although the full tensor was implemented, it proved to be unstable and noisy, such that for practical purposes the full tensor is in the rest of this work simplified to 
\[\vb*{\Pi}_h \approx p_\parallel \vb{b} \otimes \vb{b}+ (\vb{I}-\vb{b}\otimes \vb{b})p_\perp\]
with 
\begin{align*}&p_\perp =\frac{m_h}{2}\int|\vb{v}\times\vb{b}|^2 f_h \Diff3 \vb{v}, 
&p_\parallel=m_h\int v_\parallel^2 f_h \Diff3 \vb{v} 
\end{align*}
This tensor is calculated using equation \eqref{eq:fullf} as
\[\vb*{\Pi}_h = m_h\sum_{i=1}^{N_p}w_i\delta[\vb{r}-\vb{r}_i] \left(v_\parallel^2 \vb{b}_i\otimes \vb{b}_i+\frac{1}{2}|\vb{v}_i\times\vb{b}_i|^2 (\vb{I}-\vb{b}_i\otimes\vb{b}_i)\right)\]
where $\vb{b}_i$ is the magnetic field at the location of marker particle $i$. The pressure tensor is then projected onto the fluid finite elements using smoothing terms to mitigate noise that are always chosen small enough not to affect physical results.  For details of this projection procedure, see Refs. \onlinecite{van2019nonlinear,hoelzl2021jorek}, while the effect of the smoothing terms and their impact on the physical results is investigated in Ref. \onlinecite{dvornova2021hybrid}. This projection is averaged over several orbits (commonly the MHD timestep) to remove high-frequency noise and then entered as an explicit momentum source term into the bulk fluid timestepping. The bulk MHD fluid timestepping remains implicit, it is only the contribution of the EPs that is considered explicitly. This is a common strategy across different hybrid codes\cite{liu2022hybrid, fogaccia2016linear, briguglio1995hybrid}. This coupling has been implemented in both the full\cite{pamela2020extended} and reduced MHD models in JOREK. 
\section{\label{sec:phase_space} Diagnostics in phase space}
To analyse details of the EP dynamics, a versatile diagnostic has been developed which can provide insights regarding real space as well as velocity space dynamics. Similar kind of diagnostics exist in other codes\cite{brochard2020nonlinear,briguglio2014analysis,wang2013hole, kim2008impact}. Consider some single-particle quantity $g_i$. E.g., the density by choosing $g_i=1$, or the energy transfer between bulk and EPs by using $g_i=\Delta E_i$, where $\Delta E_i$ denotes the change of particle kinetic plus potential energy. The distribution of $g_i$ throughout the real and velocity space is given by the following expression:
\[\sum_{i=1}^{N_p}g_i w_i \delta[\vb{r}-\vb{r}_i(t)]\delta[\vb{v}-\vb{v}_i(t)].\] 
Although this yields all information about the distribution of $g_i$, it is an inconvenient representation. Some coordinates might not be of interest (e.g. gyrophase or toroidal angle) and some coordinates are not natural coordinates for particles moving in electromagnetic fields (e.g. $\vb{v}$ is inconvenient compared to coordinates like the magnetic moment $\mu$ and parallel velocity $v_\parallel)$.  Therefore, the idea is to apply a coordinate transformation and integrate over coordinates that are ignorable for the considered analysis. Thus, to obtain the distribution $G$ of $g_i$ in the coordinates $\vb*{\xi}$, this yields 
\begin{equation}G(\vb*{\xi},t) = \sum_{i=1}^{N_p} g_i w_i \delta(\vb*{\xi}-\vb*{\xi}_i(t)).\end{equation}
An example usage is the investigation of resonances by visualising the distribution of energy loss or gain (integrated over some time) in terms of $(v_\parallel,\mu)$, yielding 
\[G(v_\parallel,\mu) = \sum_{i=1}^{N_p} \Delta E_i w_i \delta (v_\parallel-v_{\parallel}^i)\delta(\mu -\mu_i).\]
Delta-functions are not useful for a diagnostic, as they consume a large amount of memory (for large number of particles), cannot be represented graphically, and are noisy. To solve these issues while keeping the useful integration properties of the delta-function, the diagnostic replaces the delta-function by finite-support kernels. These finite-support kernels $K_l$ are defined by 
\begin{align*}
    \int_{\mathbb{R}} K_l(x) \diff x = 1 \\ 
    K_l(x) = 0 \text{ if } |x| > l
\end{align*}
where $l$ is the so-called bandwidth. This controls the smallest visible structures and thus the amount of smoothing. Denote the amount of dimensions of $\vb*{\xi}$ by $\alpha$. The $\alpha$-dimensional kernel $K^\alpha(\vb*{\xi})$ can then be written as 
\[K^\alpha_{\vb{l}}(\vb*{\xi}) = \prod_{j=1}^\alpha K_{l_j}(\xi^j),\]
where $\vb{l}=\{l_j\}$. This multi-dimensional kernel still has the property
\[\int_{\mathbb{R}^\alpha} K^\alpha_{\vb{l}}(\vb*{\xi})\Diff\alpha\vb*{\xi}=1.\]
Then, by substitution with $\delta(\vb*{\xi}-\vb*{\xi}) \to K^\alpha_{\vb{l}} (\vb*{\xi})$ the diagnostic implemented in JOREK is obtained: \medbreak
\begin{equation}\label{eq:G_jorek}G_{\text{J}}(\vb*{\xi},t)=\sum_{i=1}^{N_p} g_i w_i  K^\alpha_{\vb{l}}(\vb*{\xi}-\vb*{\xi}_i(t)). \end{equation}
Numerically, an $\alpha$-dimensional grid is constructed. For every marker particle $i$ its contribution is added at every position $\vb*{\xi}$ where the kernel is non-zero. For some applications (such as resonance visualisation), it is useful to integrate the projection over time to reduce noise, which in this case means adding contributions of many timesteps corresponding to many gyro-motions.

Applications include visualising the full distribution function in various coordinates, e.g., in terms of $(R,Z,E,\lambda=v_\parallel/v)$ or $(E,\mu,P_\phi)$, visualising resonances as a function of various coordinates, e.g. in terms of $(v_\parallel,\mu)$ or $(E,P_\phi)$,  visualising resonant particle density as a function of the minor radius $r$ or poloidal magnetic flux $\psi$ to probe EP transport or saturation mechanisms, etc. For several figures in this work this diagnostic is already used  (e.g. figures \ref{fig:aug_phase_space}, \ref{fig:itpa_relaxation}, \ref{fig:density_flatten}, \ref{fig:itpa_relaxation}).

For resonance visualisation, a complicating factor is that during a gyro-motion the particle may lose and gain energy while there is no net energy transfer over the full gyromotion. Resonant particles mainly lose net energy due to the interaction of the drift velocities with EP instabilities\cite{heidbrink2008basic}. This net energy loss can be much less than the magnitude of the oscillation during the gyro-motion. As for a full-orbit particle, $(v_\parallel,\mu)$ are oscillating during the gyro-motion, the energy loss or gain is deposited at slightly different coordinates. This leads to the actual resonances potentially being obscured by an irrelevant oscillation (which might still be physical and not noise; e.g. particles all gain and lose energy at the same phase-space locations). The gyro-motion of the particle is then fitted to obtain only the trend of $v_\parallel$ and $\mu$, without the oscillation, such that the energy loss and gain during the gyromotion cancels. This procedure is shown in detail in Appendix \ref{app:least_squares}.  However, for resonance visualisation in coordinates that are conserved for full-orbit particles (such as $E,P_\phi$), this is not necessary.

\section{\label{sec:benchmarks} Benchmarks}
In this section, two separate benchmarks are considered. The first is the well-known ITPA case\cite{konies2008kinetic} concerning a TAE in a high aspect ratio tokamak. The second is the AUG-NLED\cite{augnled,vlad2021linear} case, which is based on a high EP pressure discharge in the realistic geometry of an ASDEX Upgrade\cite{kallenbach2017overview} tokamak discharge. The results are compared to other codes in terms of mode structure, frequency, growth rates and phase space resonances. 
\subsection{\label{sec:itpa}ITPA case}
The ITPA case and the results from other codes are described in Ref. \onlinecite{konies2008kinetic}. It concerns a high aspect ratio tokamak (major radius $R=10$ m, minor radius $a=1$ m), with a flat hydrogen bulk fluid density ($n_b = 2\cdot 10^{19}$ m$^{-3}$). The $q$-profile is $q=1.71+0.16(r/a)^2$, such that an $n=6$, $m=10,11$ TAE gap is expected at $r=0.5a$. The EPs follow a Maxwellian temperature distribution with varying fast particle temperature $T_f$. The simulation is restricted to a single toroidal mode number $n=6$. This benchmark was previously performed in JOREK for the reduced MHD model using isotropic pressure coupling in Ref. \onlinecite{dvornova2021hybrid}, but here the full MHD model is used with the anisotropic pressure coupling scheme. \medbreak
The MHD timestep was chosen as $3 \tau_A$ where $\tau_A\equiv a \sqrt{\mu_0 \rho_0} /B$ is the Alfvén time ($\rho_0$ is central mass density, $B$ is central magnetic field and $a$ is the minor radius) while the particle timestep was chosen as $2\cdot 10^{-10} \text{ s }\approx 200 \tau_g$ where $\tau_g$ is the time it takes for the EPs to complete one gyromotion. The grid was chosen to be 102 radial elements with 128 poloidal elements. 40 million particles were used.

\begin{figure*}
    \centering
    \includegraphics[width=\linewidth]{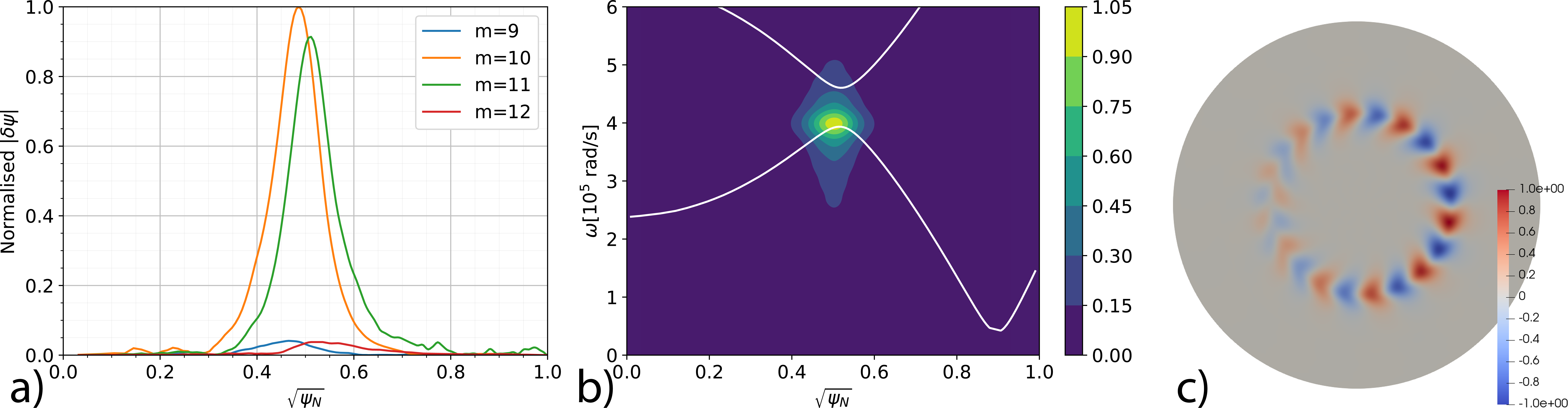}
    \caption{The poloidal harmonic structure of $\delta \psi$ (a), normalised frequency spectrum of $|\delta \psi| $ (b) and normalised poloidal plane structure of $\delta \psi$ (c) of the $n=6$ toroidal harmonic perturbation of $\psi$ in the ITPA benchmark for a simulation with $T_f=400$ keV. The white lines in (b) denote Alfv\'{e}n continua calculated using the HELENA\cite{huysmans1991isoparametric} and CASTOR\cite{kerner1998castor} codes}
    \label{fig:itpa_representative}
\end{figure*}

\begin{figure*}
    \centering
    \includegraphics[width=\linewidth]{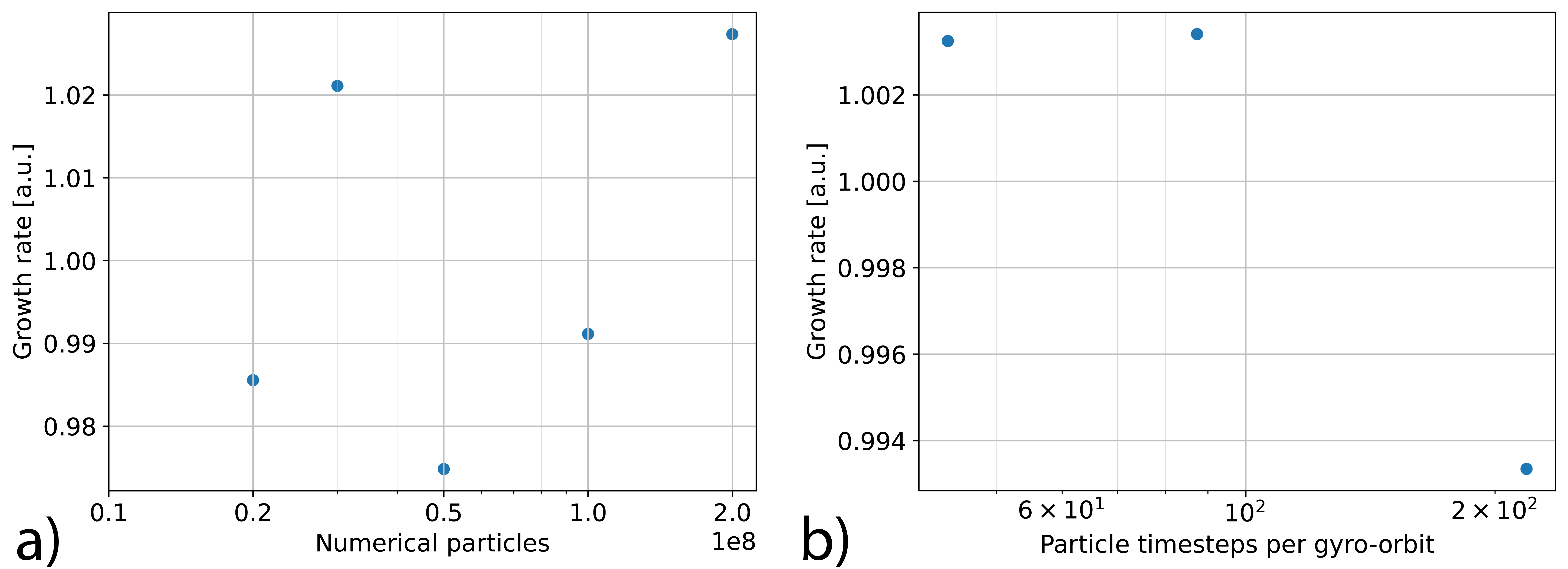}
    \caption{Normalised (to the mean) growth rate for the ITPA-TAE case with $T_f=700$ keV for varying numerical particles (a) and varying particle timesteps (b), showing that all results agree within 5\%. }
    \label{fig:convergence}
\end{figure*}

\begin{figure*}
    \centering
    \includegraphics[width=\linewidth]{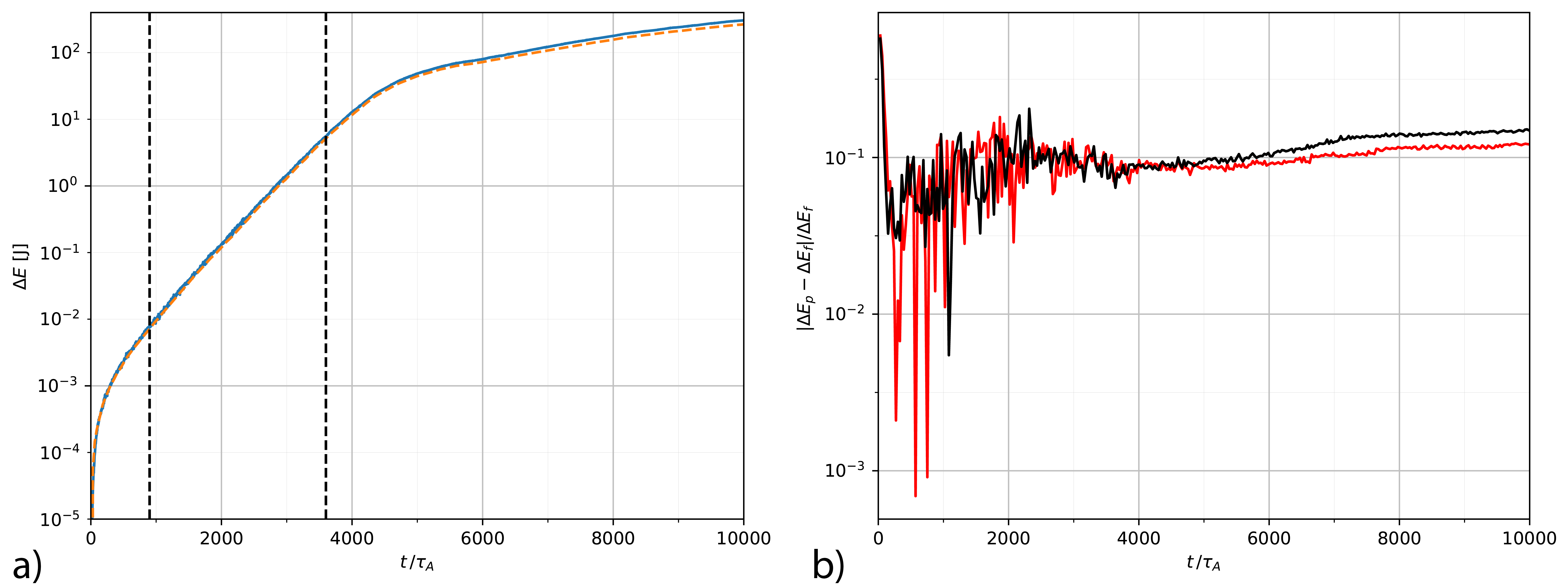}
    \caption{ (a): The energy lost by the EPs (blue solid lines) and the energy gained by the bulk fluid (orange dashed lines) as a function of time in the ITPA-TAE case with $T_f=400$ keV. During the linear phase (the region between the black dashed lines), energy is conserved well. In the saturation phase, the missing dissipation due to a fixed equlibrium leads to energy not being conserved as well. (b): The difference between fluid energy gained and particle energy lost, normalised by the fluid energy gained. The black line is the simulation in (a), while the red line is a simulation with resistivity, viscosity and numerical resistivity and viscosity reduced by an order of magnitude, showing that the loss in energy conservation in the non-linear phase is (at least partially) caused by a fixed equilibrium.  }
    \label{fig:energy_conservation}
\end{figure*}

The poloidal mode structure and frequency for a simulation with $T_f=400$ keV are shown in figure \ref{fig:itpa_representative}, exhibiting a clear TAE structure. The energy in the $n=6$ harmonic is shown in figure \ref{fig:energy_conservation}, where also the linear phase is indicated. 
A simulation set-up using $T_f =700$ keV was repeated for varying particle timestep and varying particle number to ensure sufficient convergence. These results are shown in figure \ref{fig:convergence}, where all results agree quantitatively within 5\%.

\begin{figure*}
    \centering
    \includegraphics[width=\linewidth]{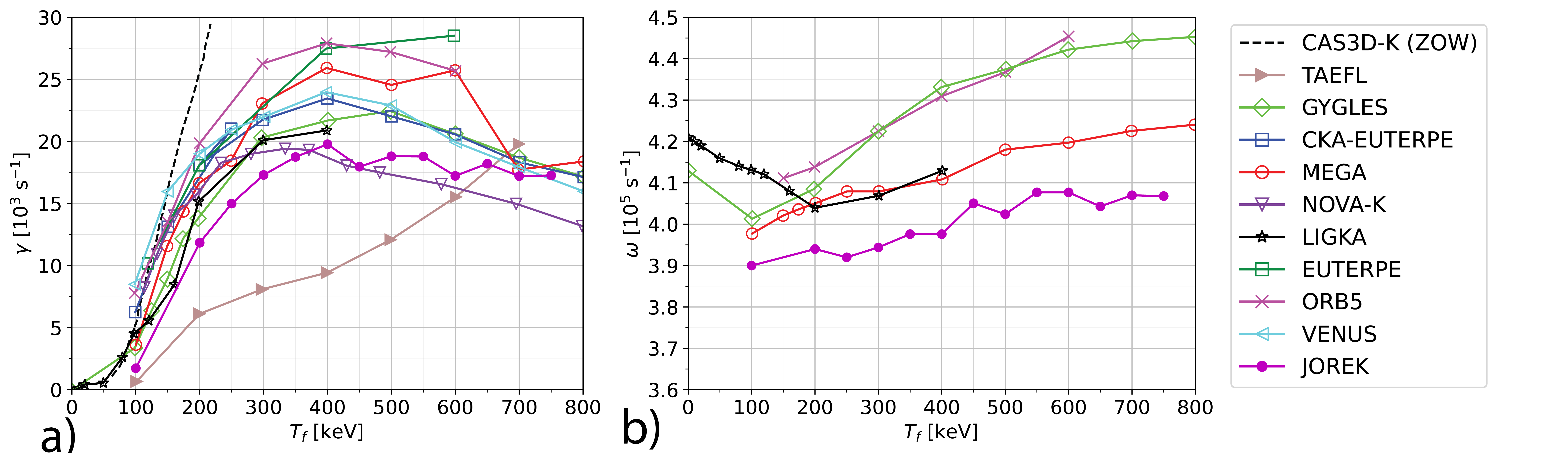}
    \caption{Growth rates (a) and mode frequencies (b) as a function of $T_f$ for the JOREK simulations compared with the results from other codes\cite{konies2018benchmark,konies2012benchmark}. Only codes which take into account finite Larmor radius effects are included in these figures.}
    \label{fig:itpa_gr_freq}
\end{figure*}

\begin{figure*}
    \centering
    \includegraphics[width=\linewidth]{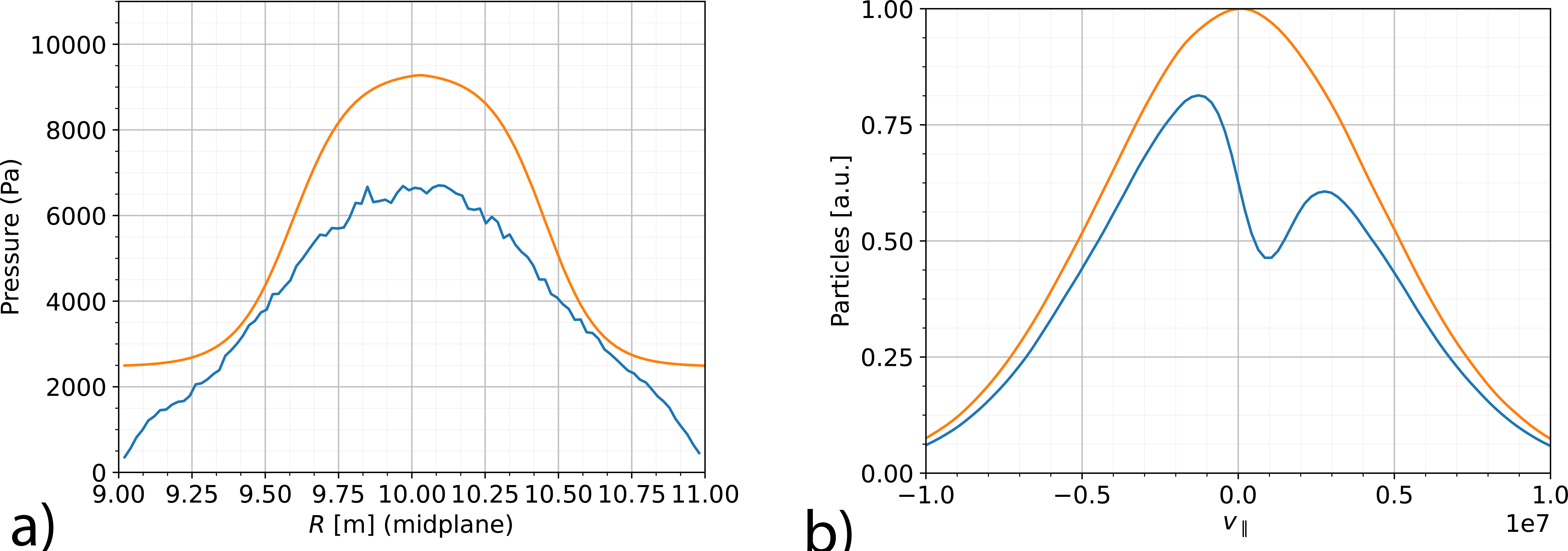}
    \caption{Midplane fast particle pressure (a) and global velocity space (in terms of $v_\parallel$) distribution (b) in the initial distribution (orange) and the distribution in the linear phase (blue) in the ITPA-TAE benchmark at $T_f=400$ keV showing a relaxation in both real and velocity space. }
    \label{fig:itpa_relaxation}
\end{figure*}

\begin{figure}
    \centering
        \includegraphics[width=\linewidth]{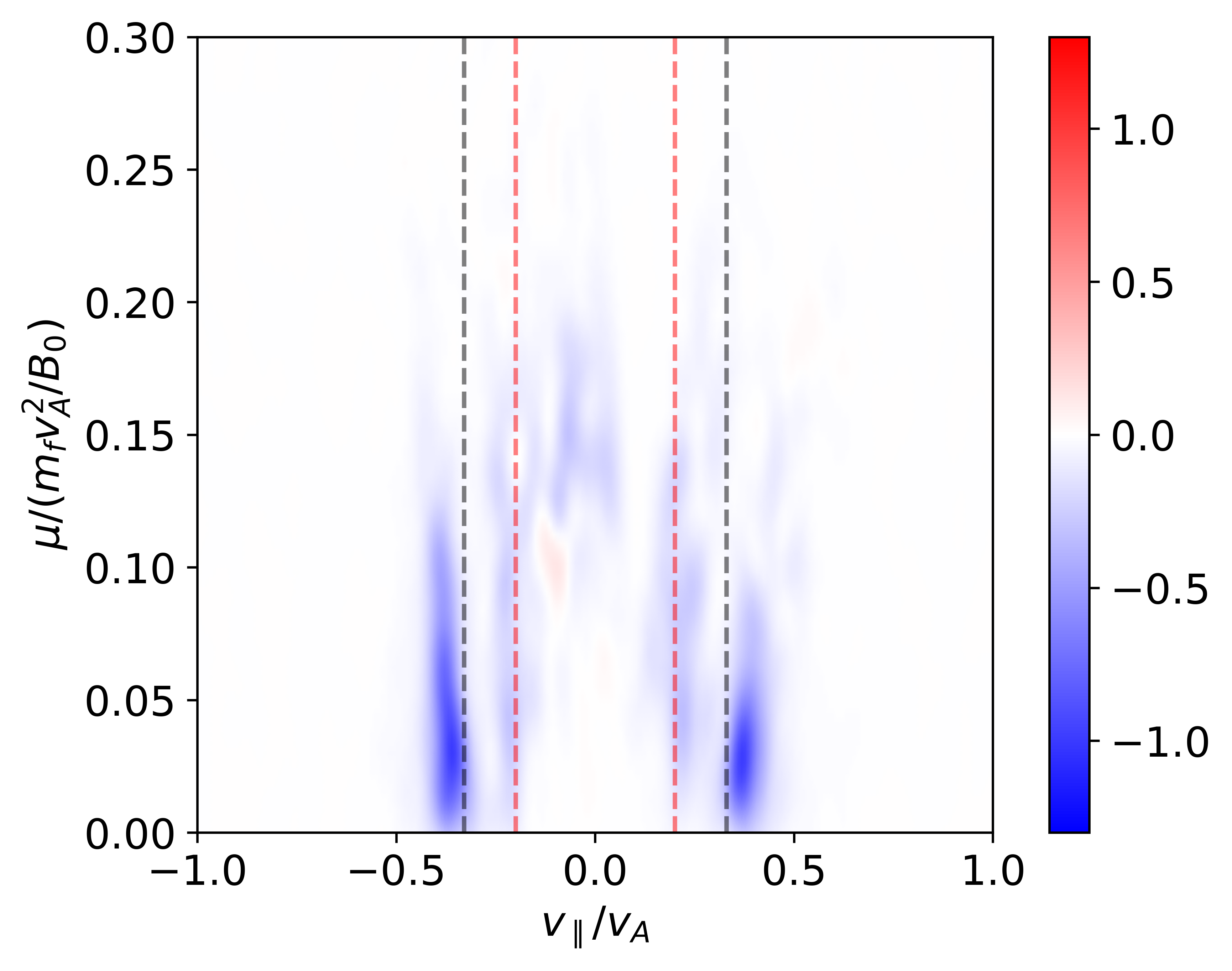}
    \caption{The normalised power gain and loss of the EPs as a function of normalised $v_\parallel$ and $\mu$ in a simulation of the ITPA-TAE case with $T_f=400$ keV during the linear phase. Dashed lines indicate theoretical resonances at $v_A/3$ (black) and $v_A/5$ (red). }
    \label{fig:itpa_phase_space}
\end{figure}

Quantitative growth-rate and frequency comparisons are given in figure \ref{fig:itpa_gr_freq}, showing reasonable agreement.
 In the other codes, the relaxation of the EP distribution function has been switched off, while this relaxation is intrincially present in JOREK due to the full-f scheme. The initialised Maxwellian distribution function relaxes as the distribution function cannot be expressed solely in terms of conserved particle quantities, changing the real and velocity space distribution functions and thus the EP drive. This relaxation is visualised for $T_f =400 $ keV in \ref{fig:itpa_relaxation}. Therefore it is difficult to compare results one-to-one in this benchmark, motivating future comparisons with fully stationary distribution functions.
In figure \ref{fig:itpa_phase_space}, the phase space resonances are visualised, showing that the $v_A/3$ resonance is dominant, in agreement with results in Ref. \onlinecite{konies2018benchmark}. These theoretical resonances are derived using assumptions on the aspect ratio (for details, see Ref. \cite{heidbrink2008basic}) and therefore do not hold exactly.

\begin{figure}
    \centering
    \includegraphics[width=\linewidth]{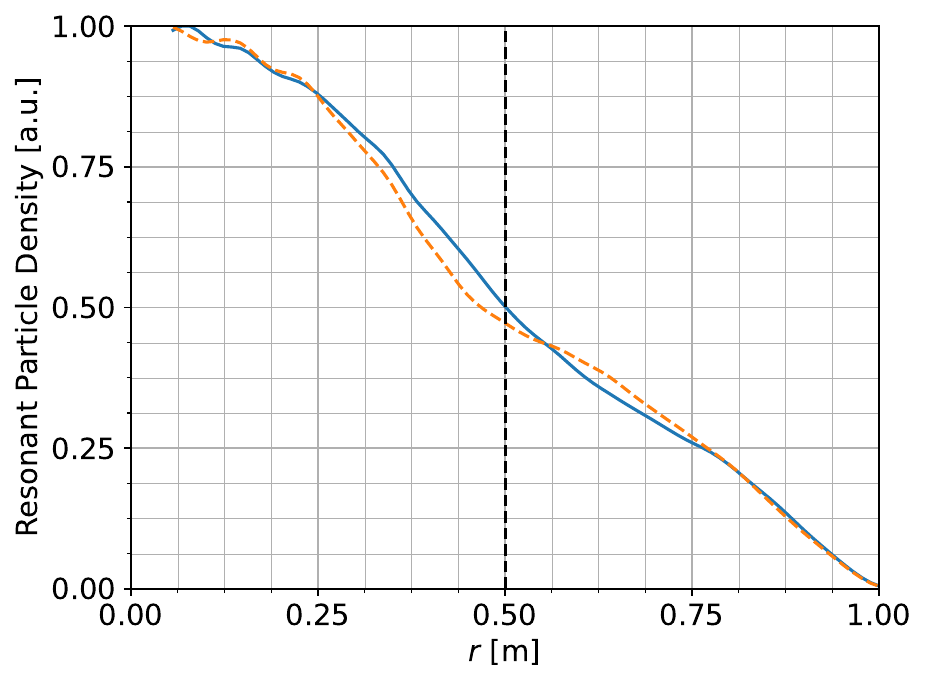}
    \caption{The normalised resonant particle density as a function of minor radius $r$ in the linear phase (blue solid lines) and during the saturation phase (orange dashed lines), showing density flattening. The black dashed line indicates the TAE gap location at $r=0.5a$.}
    \label{fig:density_flatten}
\end{figure}

For simple non-linear behaviour, the 400 keV simulation has been continued into the non-linear phase. Using the results from figure \ref{fig:itpa_phase_space}, resonant particles are selected, and their density before and after saturation is shown in figure \ref{fig:density_flatten}. Density flattening of the resonant particle density is observed in the vicinity of the mode location, as expected. 
Although the pressure coupling scheme does not conserve energy exactly, figure \ref{fig:energy_conservation} showns that energy is conserved rather well until the saturation phase. This simulation was performed with fixed equilibrium  ($n=0$), and non-zero resistivity and viscosity. Therefore, the energy associated with the dissipation of the mode is not conserved, leading to a difference between EP energy and bulk energy at the later, saturation stage, as shown in figure \ref{fig:energy_conservation}. 
\subsection{\label{sec:augnled}AUG-NLED case}
\begin{figure*}
    \centering
    \includegraphics[width=\linewidth]{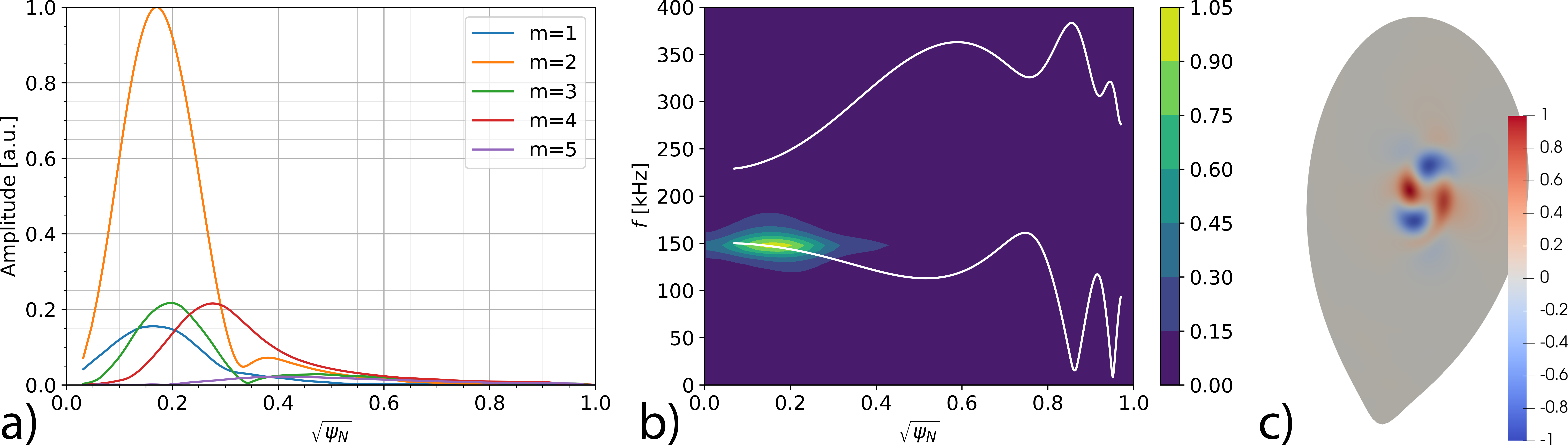}
    \caption{TThe poloidal harmonic structure of $\delta \psi$ (a), normalised frequency spectrum of $|\delta \psi| $ (b) and normalised poloidal plane structure of $\delta \psi$ (c) of the $n=1$ toroidal harmonic perturbation of $\psi$ for the nominal AUG NLED benchmark case. The white lines in (b) denote the Alfv\'{e}n continua calculated by the HELENA\cite{huysmans1991isoparametric} and CASTOR\cite{kerner1998castor} codes.}
    \label{fig:aug_represent}
\end{figure*}

The AUG-NLED case is described in Refs. \onlinecite{augnled,vlad2021linear}. It is based on an experimental AUG discharge (\#31213 at $t=0.84$s) with high EP pressure. Here, only the case with off-axis peaking of the EP distribution is considered and the reduced MHD model is used as results agree with full MHD while computational costs are lower. The EP distribution is assumed to be an isotropic Maxwellian at $T=93$ keV, while the EP density is varied. The poloidal mode structure and frequency spectra for the nominal case (as described in Ref. \cite{vlad2021linear}) is shown in figure \ref{fig:aug_represent}. From the mode structure and the mode location in the continuum, this mode can be identified as an EPM.
The Maxwellian initial distribution is not stationary as the distribution function cannot be expressed in terms fo conserved particle quantities. Thus, the distribution function undergoes a fast relaxation in the full-f treatment applied here. This case in realistic geometry relaxes much more than the ITPA case such that results cannot be directly compared anymore without compensating for the realxation in some way. In order to still be able to compare growth rates and frequencies, the pressure gradient after relaxation was varied using two different methods. First, by simply increasing the weight of the particles and thus scaling inital EP total density, and secondly by modifying the initial profile such that the gradient at the mode location after relaxation is higher. The equilibrium (and thus the Alfv\'{e}n continuum) is not changed in this procedure though (as increasing the density artificially would lead to a very low bulk mass density). 

In all cases, 10 million particles were used. 51 radial elements were used with 64 poloidal elements. Timesteps of $0.26  \ \mu$s$\ \approx 4 \tau_A$ were used for the MHD fluid and $2\cdot 10^{-10}$ s for the EPs.

\begin{figure*}
\centering
    \includegraphics[width=\linewidth]{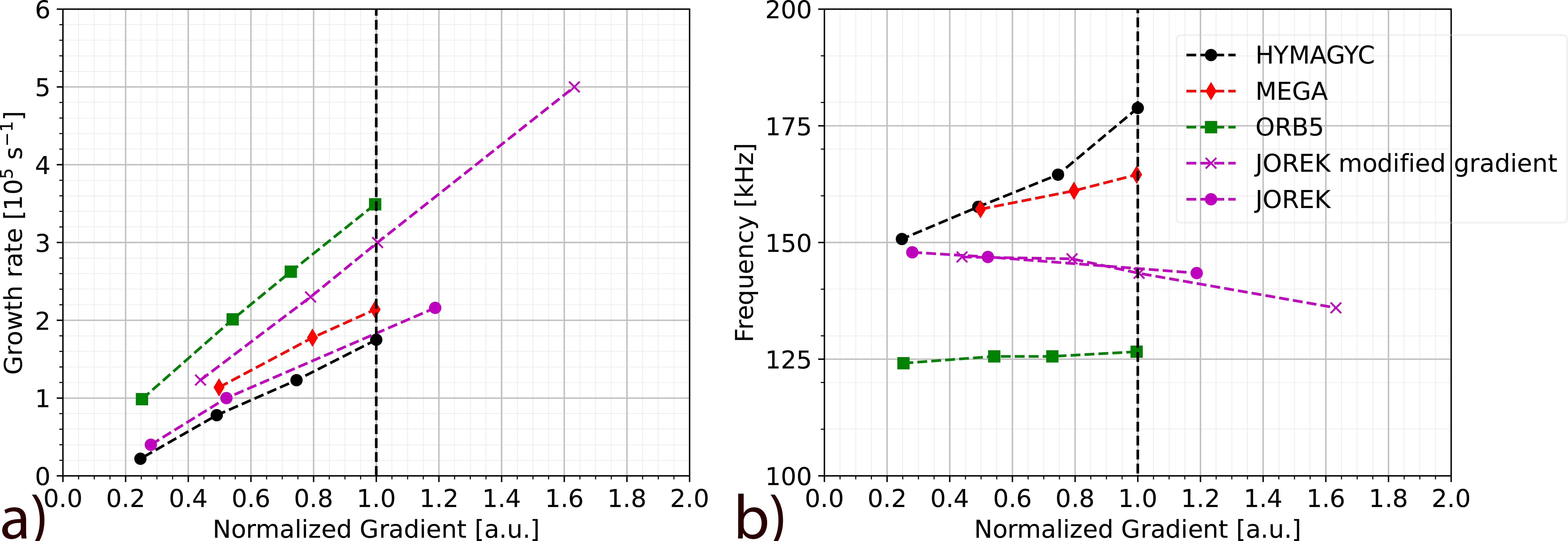}
    \caption{Growth rates (a) and mode frequencies (b) as a function of the normalised pressure gradient in the AUG NLED case for JOREK and other codes\cite{vlad2021linear}.}\label{fig:AUG_gr_freq}
\end{figure*}

The comparison with other codes is shown in figure \ref{fig:AUG_gr_freq}. As in the ITPA case, the comparison is not directly one-to-one anymore due to the real and velocity space relaxation, but the agreement is good in terms of the dependency of EP drive on the presssure gradient, in terms of the mode structure excited, and in terms of the mode location in the continuum. The frequency does not match as closely. In the original benchmark results the bulk ion density was different for different EP densities while the bulk ion density was constant in the JOREK runs (as noted above). Therefore, the Alfv\'{e}n continua in the JOREK runs will be slightly different compared to the original benchmark results. The  frequencies of the waves that can be excited will thus differ as well, providing an explanation for the difference in frequencies between JOREK and the original results for higher EP densities. 

\begin{figure*}
    \centering
    \includegraphics[width=\linewidth]{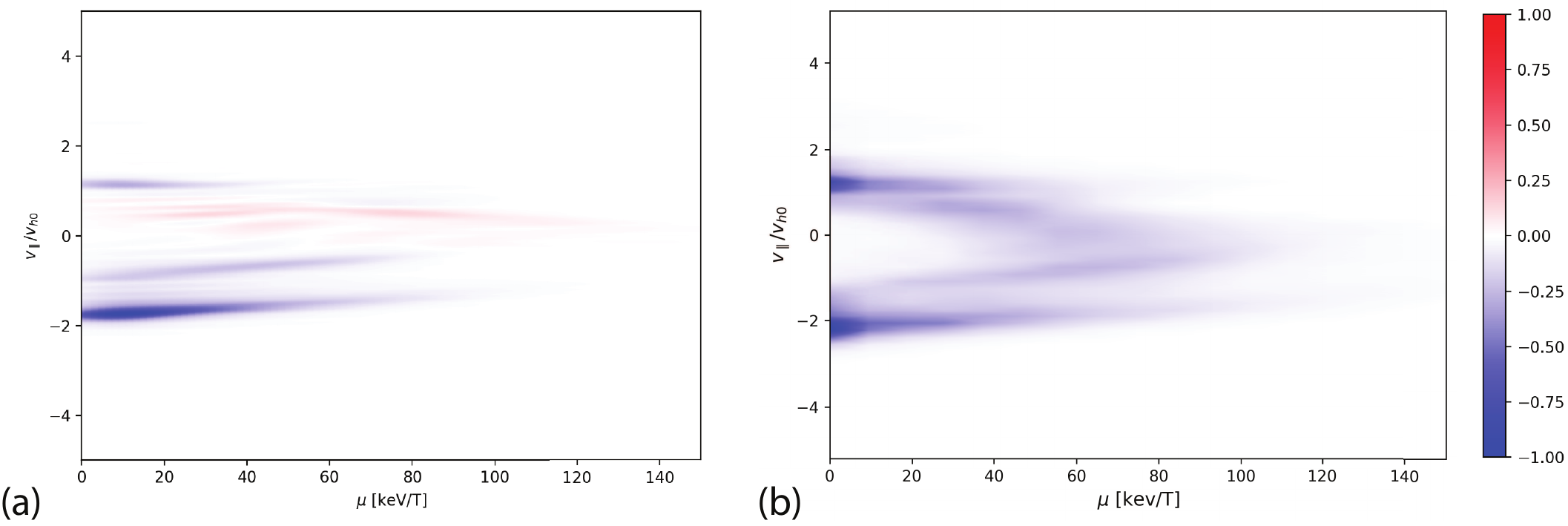}
    \caption{Phase-space resonances, i.e., energy transfer between particles and bulk plasma in terms of $v_\parallel$ and $\mu$ in JOREK (a) and MEGA (b) for the nominal AUG NLED case. The parallel velocity is normalised to the thermal EP velocity as $v_{h0}=\sqrt{E_h/m}$. Both codes show a dominant resonance at $v_\parallel/v_{h0}\approx -2$, and no resonance at $v_\parallel/v_{h0}\approx 2$.  }
    \label{fig:aug_phase_space}
\end{figure*}

Phase space resonances are also compared with MEGA for the nominal case in figure \ref{fig:aug_phase_space}, showing similar features but quantatitive differences. These differences can arise due to the very different models (full-$f$ and full-orbit in JOREK compared to $\delta f$ and drift-kinetic that was used in MEGA), and the relaxation of the distribution function in real and velocity space. Furthermore, note from figure \ref{fig:AUG_gr_freq} that MEGA finds a frequency near 165 kHz in the nominal case while JOREK (without correcting for gradient) obtains about 150 kHz, which also impacts resonances. As EPMs, even in a linear description, require a nonperturbative analysis of the EP response\cite{chen1994theory}, these differences in the distribution function and the frequency of the mode could explain the quantitative differences. 

\section{\label{sec:application} Application to an AUG discharge}
\begin{figure}
    \centering
    \includegraphics[width=\linewidth]{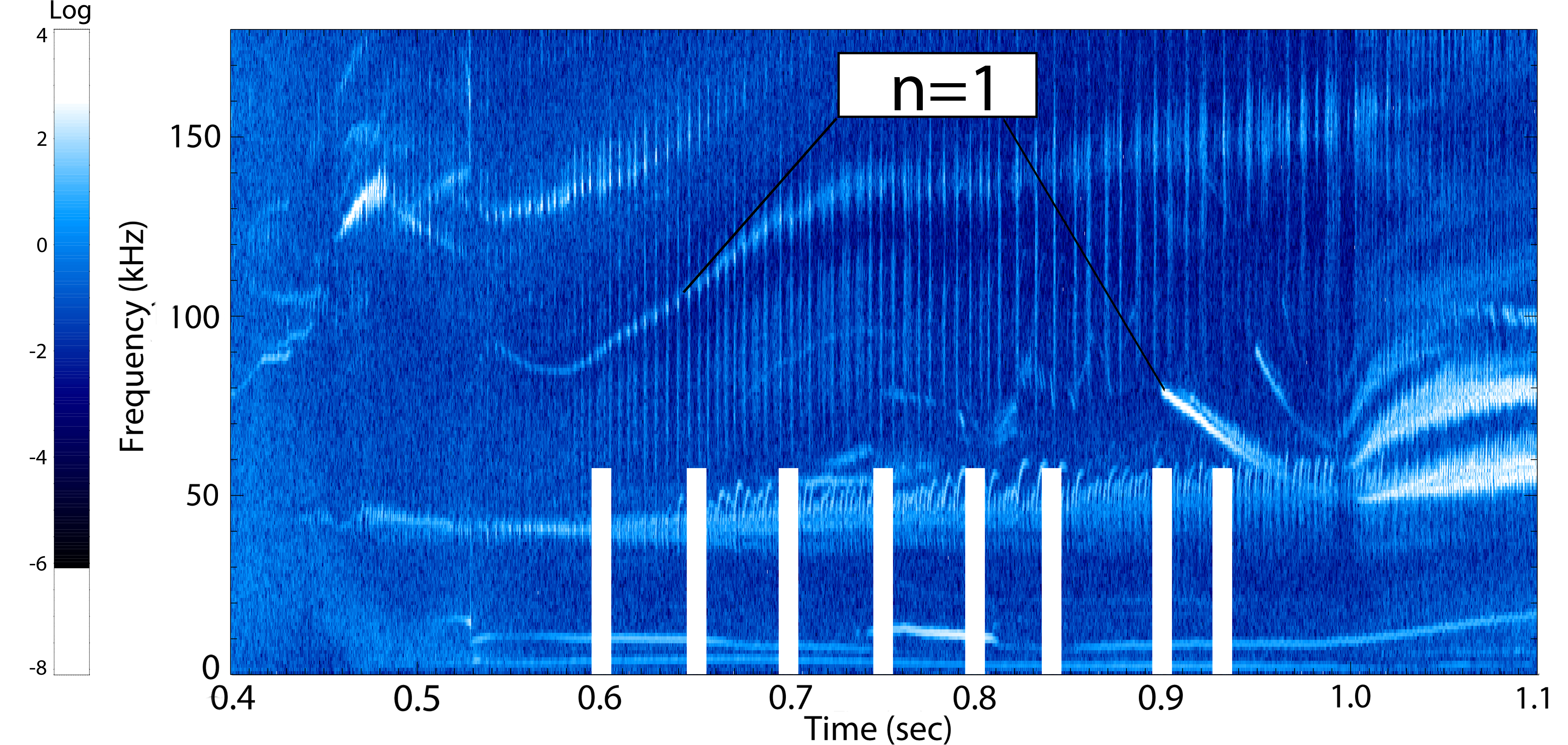}
    \caption{Spectrogram obtained from Mirnov coil measurements for the AUG-NLED discharge. The white lines denote times considered in this work. The highest amplitude $n=1$ toroidal harmonic modes are indicated with the black lines. Intensity of the signal is indicated by the color. The low-frequency modes at around 50 kHz (from 0.4s to 1.0s) are EGAMs \cite{vannini2021gyrokinetic}. The modes at around 100 to 150 kHz indicated by the arrow (rising from 0.5s to 1.0s) are $n=1$ Alfvénic modes considered here. The 50 to 70 kHz mode indicated by the white arrow was identified as a $n=1$ RSAE. Spectrogram courtesy of Philipp Lauber. }
    \label{fig:spectrogram}
\end{figure}

To validate the code with realistic parameters, it is applied to a high EP pressure discharge using experimental plasma profiles and a realistic EP distribution function. The AUG discharge considered is, as in the AUG NLED benchmark, discharge \#31213. The spectrogram is shown in figure \ref{fig:spectrogram}. The goal of this section is to reproduce some characteristics of the spectrum, such as emergence of modes and frequency sweeping of other modes. To this end, several timepoints are simulated. These time slices are taken at $\{0.6,0.65,0.7,0.75,0.8,0.84,0.9,0.93\}$ and are indicated in figure \ref{fig:spectrogram}. The simulations are restricted to $n=1$ for simplicity.

AUG is equipped with a large number of diagnostics, which are used via an integrated data analysis (IDA) framework to obtain accurate equilibria profiles\cite{weiland2018rabbit,fischer2019sawtooth,fischer2020estimation,fischer2010integrated,fischer2016coupling}. The equilibria and plasma profiles are directly imported into JOREK for these simulations. As the spectrum is expected to consist of core-located modes, the simulation domain does not extend beyond the separatrix. The bulk ion charge density can be obtained by subtracting the fast ion (charge) density from the electron (charge) density. The bulk mass density can be obtained by dividing the ion charge density by the effective charge of the ion population and multiplying by the effective mass of the ion population. The main impurity is assumed to be Boron (atomic mass of 10-11) as in the IDA equilibrium reconstruction \cite{fischer2019sawtooth} and the rest of the plasma is deuterium, such that the effective atomic mass per unit charge is approximately two. The choice was made to treat the plasma as a fully deuterium plasma for the purposes of the bulk mass density (as in AUG NLED benchmark setup). 

 The MHD temperature $T=T_i+T_e$ is set as to reproduce the pressure of the IDA equilibrium. The $FF^\prime$ profile is imported directly. Then, the built-in Grad-Shafranov solver in JOREK is used to obtain a discretely accurate force balance in the initial conditions. These JOREK-calculated equilibria can also be used to calculate the Alfvén continua with the HELENA and CASTOR codes. The MHD temperature is not set to an experimental profile but to reproduce the equilibrium pressure, and thus temperature-dependencies in viscosity and resistivity terms are not desired. These dependencies are switched off, leading to a fixed value of the viscosity and resistivity.
 
\begin{figure*}
    \centering
    \includegraphics[width=\linewidth]{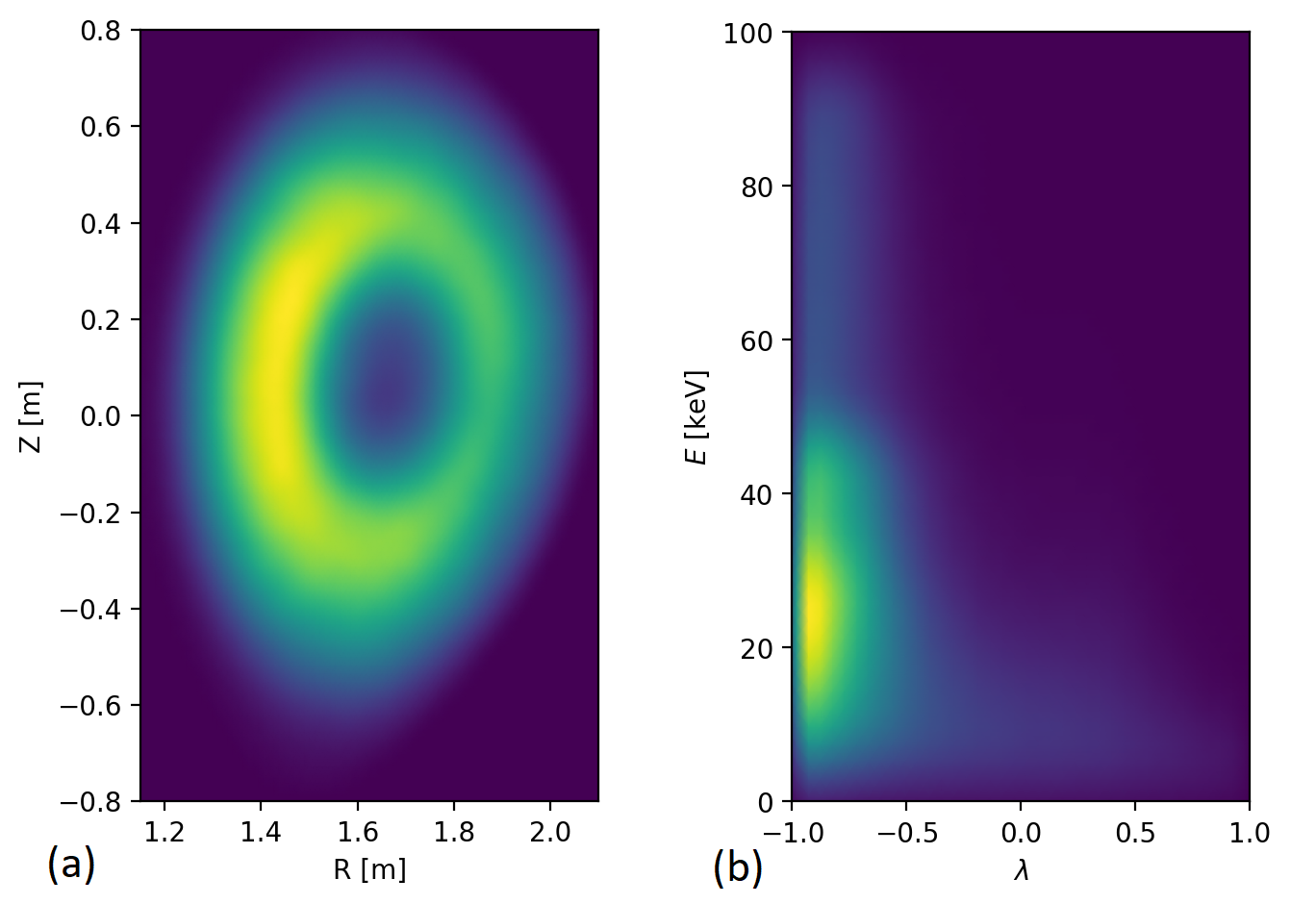}
    \caption{The spatial distribution function (a) and the velocity space distribution function (b) (at one particular position, but this does not vary significantly) of the realistic fast particle distribution function used in the simulations of AUG discharge \#31213. $\lambda$ is the pitch angle $v_\parallel/v$. }
    \label{fig:AUGD_rzproj}
\end{figure*}

\begin{figure*}
    \centering
    \includegraphics[width=\linewidth]{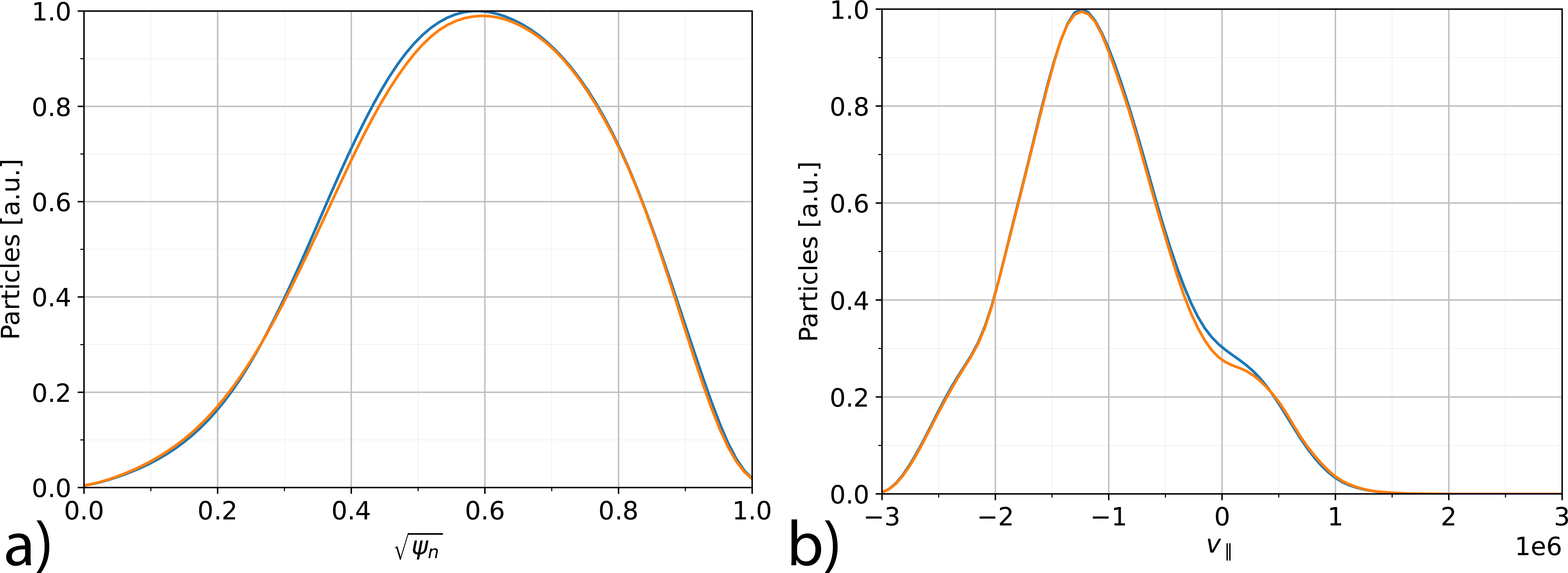}
    \caption{The normalised initial distribution function (blue) and the relaxed distribution in the linear phase (orange) in the real space (a) and in terms of $v_\parallel$ (b) of an AUG simulation using the realistic distribuiton function in figure \ref{fig:AUGD_rzproj}.}
    \label{fig:augd_evolution}
\end{figure*}

 A highly anisotropic, realistic distribution function is used, obtained from Ref. \cite{augnled}.  This distribution was calculated with the NUBEAM \cite{pankin2004tokamak} NBI code and is shown in figure \ref{fig:AUGD_rzproj}. This distribution function was found to relax much less in both velocity and real space, as can be seen in figure \ref{fig:augd_evolution}
 
 As the NBI beam was injected co-current, the particles are almost solely counterpassing (i.e. not trapped and propagating in the direction opposite to the magnetic field). A slight complication is that the EPs (in a stationary state like used here) are not uniformly distributed along the poloidal angles, such that in principle the EP density is not solely a function of $\psi$. However, this is not taken into account at present, and the flux-surface averaged density is used to obtain the bulk mass density. The EP marker particle weight is then set such that the total amount of particles is equal to the integrated flux-surface averaged density from the IDA diagnostics. 
 
 20 million particles were used for all timepoints. 51 radial elements were used with 64 poloidal elements. Timesteps of $\approx 4 \tau_A$ were used for the MHD fluid and $2\cdot 10^{-10}$ s for the EPs.
 
\begin{figure}
    \centering
    \includegraphics[width=\linewidth]{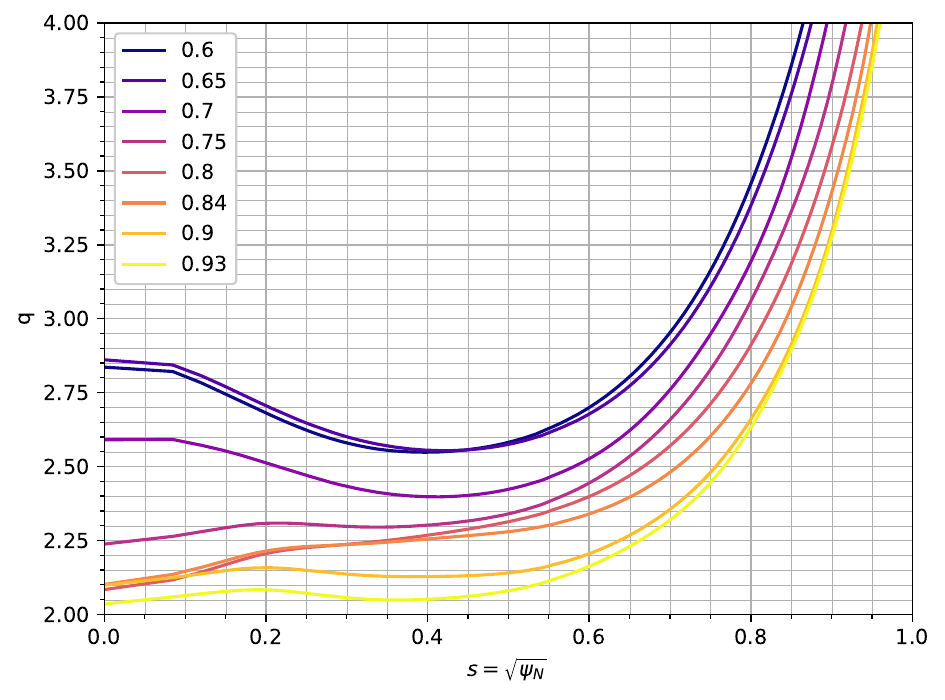}
    \caption{The $q$-profile of AUG discharge \#31213 obtained via the IDA diagnostics at the times considered in this work, showing a steady decrease in time. Furthermore, after $t=0.75$, the $q$-profile drops considerably in the core, leading to a low shear reagion at $s<0.4$.}
    \label{fig:q_aug}
\end{figure}

 Before the results are shown, it is worthwhile to shortly discuss the evolution of the discharge. The $q$-profile evolution obtained from the IDA data is shown in figure \ref{fig:q_aug}. The discharge is characterised by a reverse shear $q$-profile, which decreases in time. In the IDA diagnostic, the $q$-profile is also quite flat in the core ($s<0.4$) for $t\geq0.75$. The EP pressure (and density) increases, but the total (EP+bulk) pressure stays consistent throughout the discharge. The bulk mass density slowly decreases in time. 

\begin{figure*}
     \centering
     \includegraphics[width=\linewidth]{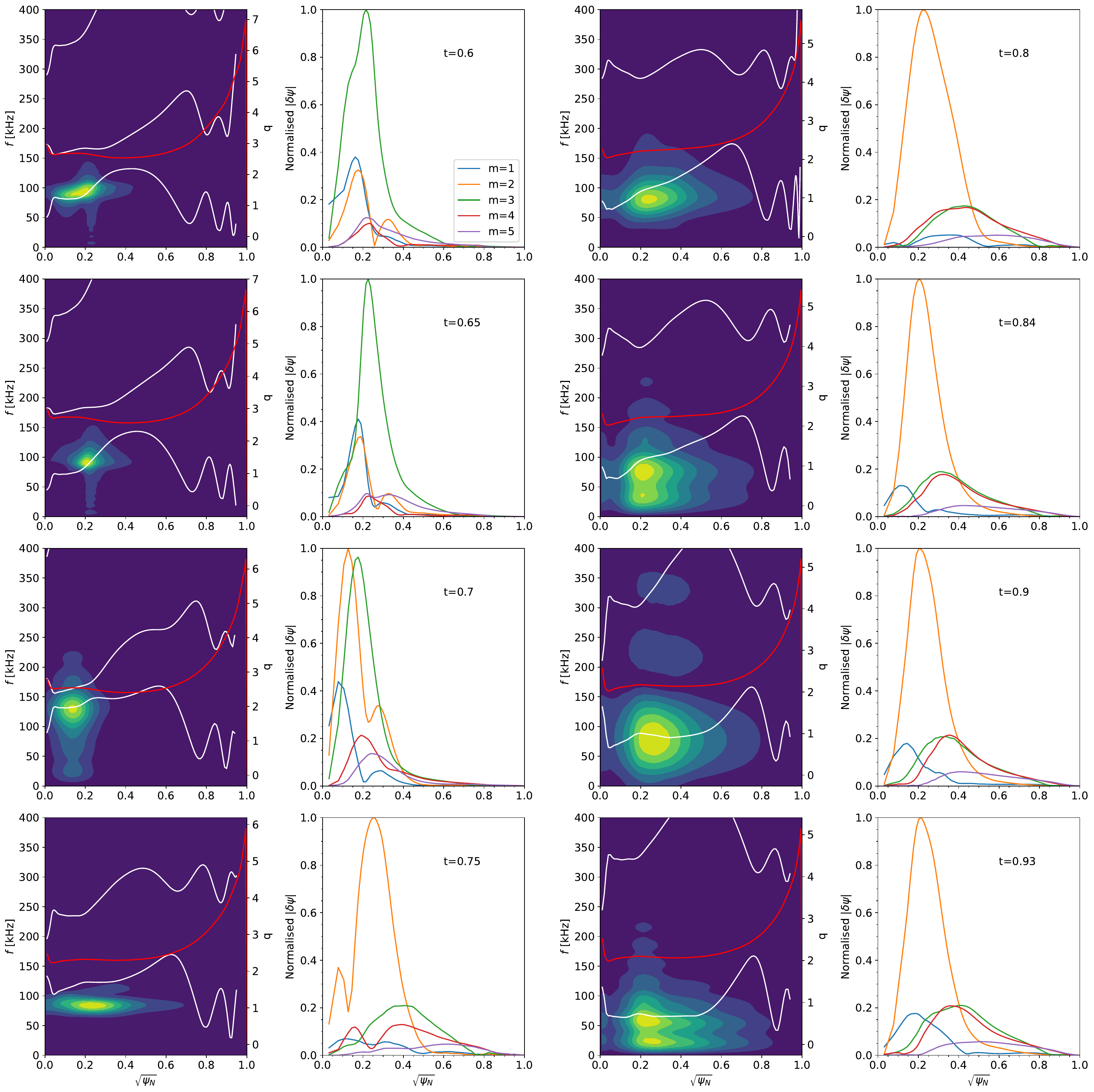}
     \caption{The frequency spectra (first and third column) and poloidal harmonic structure (second and fourth column) for the times considered in this work. The times are indicated in the poloidal harmonic structure plots. The white lines denote the Alfv\'{e}n continua calculated with the HELENA\cite{huysmans1991isoparametric} and CASTOR\cite{kerner1998castor} codes. The red lines denote the $q$-profile, such that the location of the modes can be compared to the low-shear regions at $t>0.75$s. }
    \label{fig:augd_all_time_points}
 \end{figure*}

\begin{figure}
    \centering
    \includegraphics[width=\linewidth]{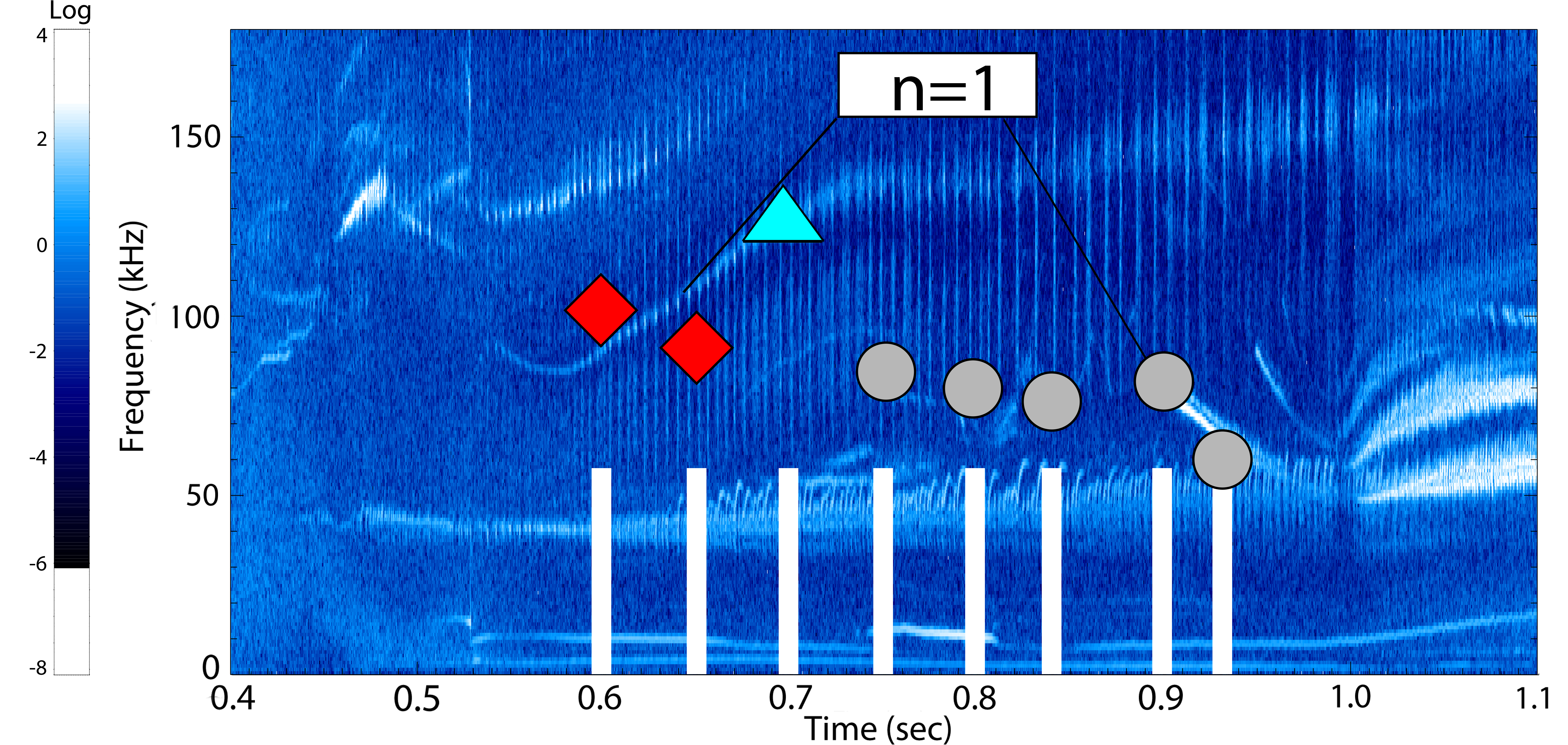}
    \caption{The spectrogram obtained from Mirnov coils during the AUG-NLED discharge including the simulated most unstable modes. The red diamonds denote the $m=3$ EPM regime modes, the cyan triangle denotes the $m=2,3$ coupled EPM mode and the grey circles denote the $m=2$ EPMs or low-shear modes. The frequencies of the most unstable modes at 0.6, 0.65, and 0.7 are a reasonable match to the spectogram. At 0.75, 0.8 and 0.84 seconds the most unstable modes do not match the spectrogram. Finally, at 0.9 and 0.93s seconds the most unstable modes match the spectogram closely. Spectrogram courtesy of Philipp Lauber.}
    \label{fig:augd_modes}
\end{figure}

\begin{figure*}
    \centering
    \includegraphics[width=\linewidth]{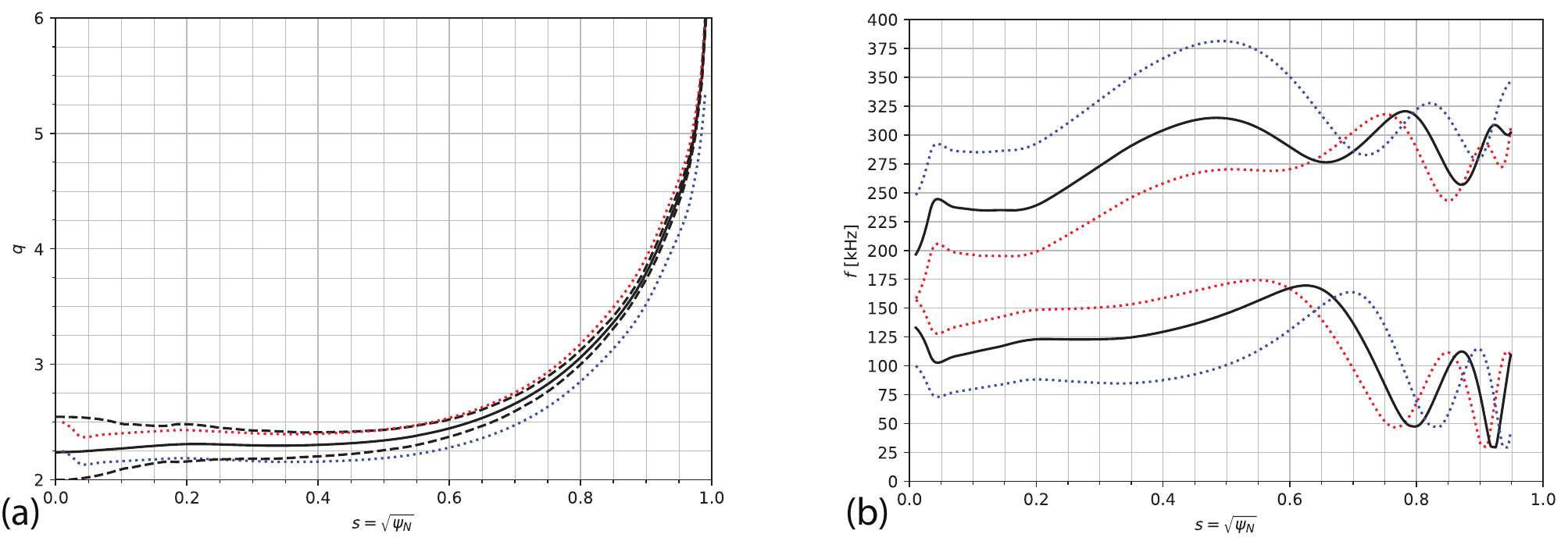}
    \caption{(a): the $q$-profile (solid black line), uncertainty in the $q$-profile (dashed black lines), and upper and lower $q$-profile used for quantifying uncertainty in the spectrum (dotted blue and red lines), all at $t=0.75$ s. (b): the Alfv\'{e}n continuum (solid black line) and the uncertainty in the spectrum (dotted blue and red lines) obtained using the blue and red dotted lines in (a). }
    \label{fig:AUGD_uncertainty}
\end{figure*}

The growth rates, frequency, total amount of EPs, and type of mode for the most unstable modes at the considered times are given in table \ref{tab:augd_gr_freq}. The frequency spectra and poloidal harmonic structure for the most unstable modes are shown in figure \ref{fig:augd_all_time_points}. It is clear that three regimes can be identified: while $q>2.5$ ($t=0.6,0.65$), the $m=3$ dominated EPMs are present. When $q$ crosses the 2.5 mark, a core TAE emerges ($t=0.7$). Finally when $q<2.5$ for $s<0.6$ ($t\geq 0.75$), the $q$-profile is flat in the core and $m=2$ dominated EPMs or Reversed Shear Alfv\'en Eigenmodes (RSAEs) emerge. 
\begin{table*}
\caption{\label{tab:augd_gr_freq}The growth rates, frequency, total amount of EPs, and type of mode for the most unstable modes found in JOREK simulations at the considered time points}
\begin{ruledtabular}
\begin{tabular}[c]{ |l|c|c|c|c|c|c|c|c| }  
 Time [s]& 0.6 & 0.65 & 0.7 & 0.75 & 0.8 & 0.84 & 0.9&0.93\\  \hline\hline
 Number of EPs [10$^{19}$] & 1.0 & 1.0 & 1.25& 1.25&  1.4 & 1.4 &  1.5   & 1.5     \\ \hline
 Frequency [kHz]  &101& 90&128 & 85&80 &76 & 82&60\\ \hline
 Growth rate [$10^5$ s$^{-1}$]& $0.5$& $0.5$&$1.0$ &$1.1$ &$1.8$ & $1.8$& $2.3$&$1.9$\\ \hline
 Type of mode & EPM & EPM & TAE &  RSAE &  RSAE &   RSAE &  RSAE &  RSAE \\ \hline
 Poloidal mode number $m$ & 3 & 3 &2, 3& 2&2&2&2&2\\
\end{tabular}
\end{ruledtabular}
\end{table*}
Superimposing the most unstable modes in the simulations on the experimental spectrogram yields figure \ref{fig:augd_modes}. Good agreement is found for $t\leq 0.7$ and $t\geq 0.9$, but the RSAEs at $\{0.75,0.8,0.84\}$ are not present in the spectrogram.
To assess the cause for this discrepancy, consider the uncertainty in the core $q$-profile. The error bars on the $q$-profile lead to uncertainties in the continua as shown in figure \ref{fig:AUGD_uncertainty}. Since the observed modes are all consistent with the Alfv\'{e}n spectra, this rather large uncertainty in the Alfv\'{e}n spectra could explain the differences. Furthermore, a modification of the core $q$-profile such that it is monotonic for $s<0.4$ could lead to these low-shear EPMs or RSAEs at $t\geq0.75$ s no longer being the most unstable modes, such that possibly modes could emerge that are visible in the experimental spectrogram.  Also, non-linear effects modifying the EP distribution function could play a role.
\section{Conclusion}\label{sec:conclusion} 
Confining energetic particles is crucial for sustaining a burning fusion reactor, but EP driven instabilities can prove a threat to this confinement. Understanding these instabilities is important for developing next-generation devices and opertional scenarios. In this work, the non-linear MHD code JOREK has been extended by anisotropic pressure coupling of EPs to the full and reduced MHD models such that it can simulate these EP instabilities using a wide variety of possible bulk physics models, powerful numerics, and grids up to the first wall. 

The JOREK model for EP simulations was explained, including a description of a versatile projection diagnostic that has been developed to investigate EP dynamics in more detail in velocity space. 

The JOREK code was then benchmarked against other codes using two cases: a TAE benchmark in simple geometry and an EPM benchmark in realistic geometry with high EP pressure. Although other codes used a different model for the EPs which does not include the relaxation of the initial Maxwellian distribution, good results have been obtained in terms of mode structure, frequency, growth rates, and phase space resonances. 

The code is then applied to a high EP pressure discharge in section \ref{sec:application} using experimental plasma geometry and parameters and a realistic distribution function for the EPs. Three regimes could be identified from these simulations and the calculated Alfv\'{e}n continua, namely an $m=3$ EPM regime, a $m=2,3$ TAE, and finally a RSAE regime. Observed frequencies do match well with the Alfv\'{e}n continua calculated from the experimental equilibria. Some of the obtained frequencies are present in the experimental spectrogram, but some are not.  The impact of uncertainties in the $q$-profile was shown to be large for the core Alfv\'{e}n continuum, providing a possible explanation for the RSAEs that are not present in the experimental spectrogram.

In the future, codes could be benchmarked in the linear and non-linear regime using realistic distribution functions and realistic or experimental plasma profiles in order to get closer to experimentally relevant scenarios. Further investigations into the AUG discharge \#31213 (possibly using multiple codes) would be of interest as well, since this discharge was designed to mimic reactor-relevant EP conditions. 

This work is only a first step regarding EP instability studies using JOREK, and not all features of JOREK have been used. For example, collisions can be used to self-consistently evolve a EP distribution function with sources and sinks, requiring, however, particular attention to energy conservation. Gyro-centre particles can be used to improve the performance after detailed comparisons to full-orbit particles. Fully consistent equilibria can be generated by evolving the equilibrium and the EPs axisymmetrically. The projection diagnostic can be used to investigate non-linear behaviour in depth. The full MHD capabilities could be used for investigating fishbone-type instabilities in realistic geometry.
\section{Acknowledgements}
 We thank Philipp Lauber for providing the spectrograms of AUG shot \#31213. This work has been carried out within the framework of the EUROfusion
Consortium, funded by the European Union via the Euratom Research and Training
Programme (Grant Agreement No 101052200 — EUROfusion). Views and opinions
expressed are however those of the author(s) only and do not necessarily reflect
those of the European Union or the European Commission. Neither the European Union
nor the European Commission can be held responsible for them.
\appendix

 \section{Optimisation of resonance visualisation diagnostic}\label{app:least_squares}
As mentioned in section \ref{sec:phase_space}, the oscillating $(v_\parallel,\mu)$ can be troubling for a resonance visualisation diagnostic. This is not a problem for resonance visualisation for quantities that are fully conserved during full-orbit motion such as the energy $E$ and the toroidal canonical momentum $P_\phi$. However, to compare with other codes and theory, it is useful to have the option to visualise resonances in terms of $(v_\parallel,\mu)$ as well. Therefore, in this appendix a method is introduced to eliminate this oscillation and illustrated with results from the ITPA benchmark of section \ref{sec:itpa}.
\begin{figure*}
     \includegraphics[width=\linewidth]{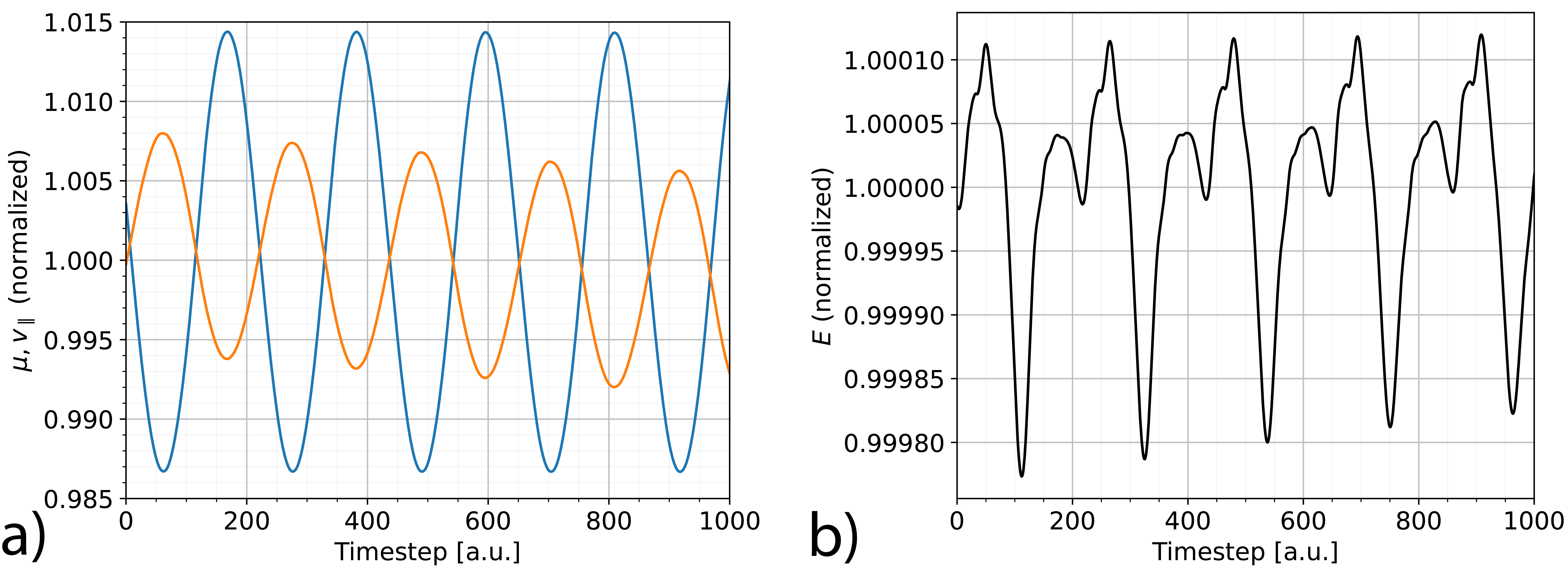}
     \caption{The normalised $\mu, v_\parallel$ (a) and the normalised energy (b)  (all normalised to the mean value) of a full-orbit particle in the ITPA-TAE equilibrium during approximately five gyro-motions, showing the oscillation of each quantitiy, together with a downwards trend for $v_\parallel$ and an upwards trend for the energy.}\label{fig:oscillation}
 \end{figure*}
During the gyro-motion the energy of a full-orbit particle varies with a larger oscillation and a smaller trend , shown in figure \ref{fig:oscillation}. 
  \begin{figure*}
      \centering
      \includegraphics[width=\linewidth]{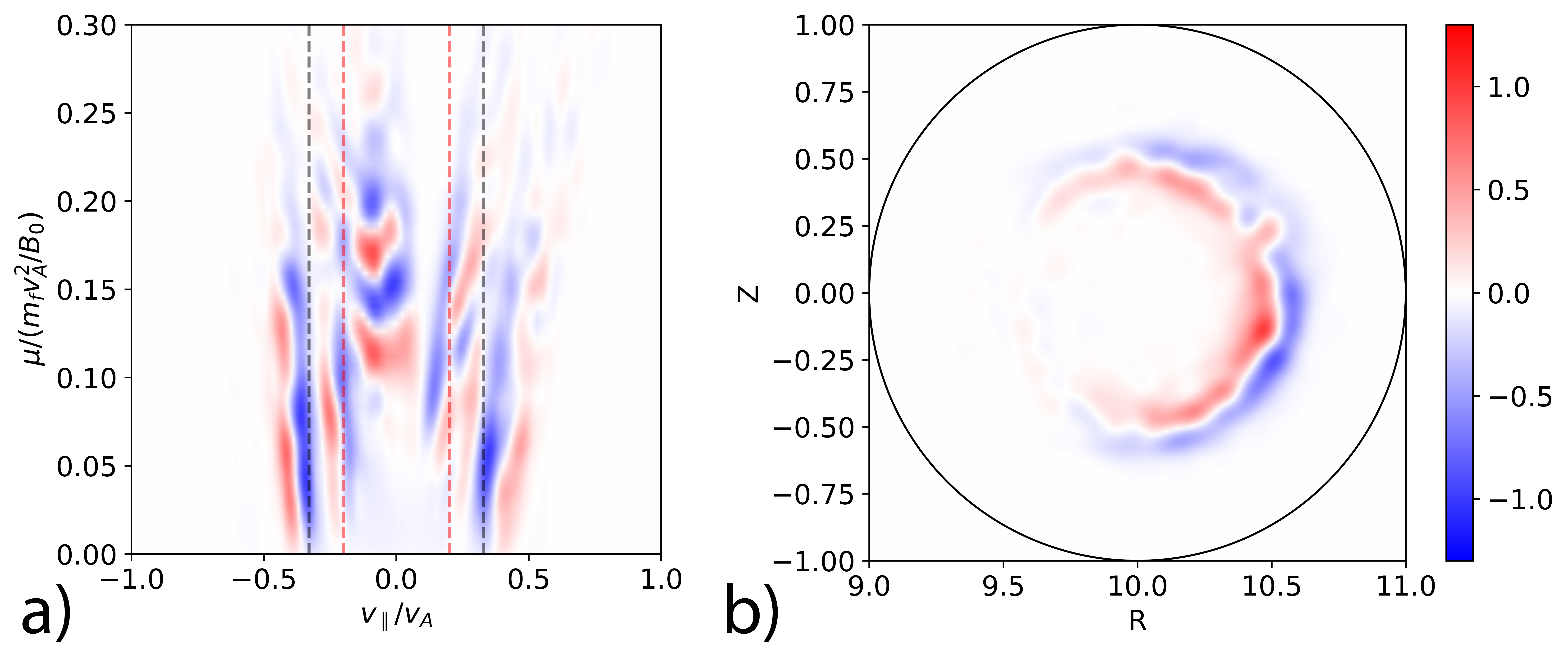}
      \caption{Energy gain or loss of the EPs in $(\mu, v_\parallel)$ space (a) and $(R,Z)$ space (b) in the ITPA-TAE benchmark. This includes the oscillation of $E$, showing dipole-like results obscuring resonances. In figure (a) the dashed lines indicate theoretical resonances, with black being $v_A/3$ and red $v_A/5$. }
      \label{fig:dipoling}
  \end{figure*}
  As the energy oscillation is of higher magnitude than the trend in the energy, the diagnostic will hide this trend in the oscillation, leading to the power exchange diagnostic producing dipole-like results in both $(R,Z)$ space and $(\mu,v_\parallel)$ space, shown in figure \ref{fig:dipoling}. Although these plots can still show the resonances (and might be interesting results in their own right), it would be more useful if the diagnostic only produces the net power exchange, without the oscillating contributions.

To convert a full-orbit particle with oscillating $v_\parallel$ and $\mu$ to a gyro-kinetic particle with fixed $\mu$ and non-oscillating $v_\parallel$, these quantities can be modeled as a linear function of time modulated with an oscillation at the gyro-frequency (within short enough timescales). The function to be fitted is then 
\[f(t,\alpha,\beta,\gamma,\delta,\omega ) =\alpha + \beta x + \gamma \cos(\omega t )+\delta \sin(\omega t).\]
Considering a set of values $\{\mu,v_\parallel \}_i = y_i$ at time $t_i$, the difference (or residual) $r_i$ between the values from the full orbit quantity $y_i$ and the function $f(t_i,\alpha,\beta,\gamma,\delta,\omega)$ at that time is 
\[r_i =y_i - \alpha - \beta t_i - \gamma \cos(\omega t_i )+\delta \sin(\omega t_i) \]
The sum of the residual squares is 
\[S = \sum_{i}r_i^2,\]
such that the partial derivative of the sum of squares by the parameter $\beta_j \in \{\alpha,\beta,\gamma,\delta,\omega\}$ is
\[\pdv{S}{\beta_j} = 2\sum_i r_i \pdv{r_i}{\beta_j}.\]
These can be calculated easily as
\begin{align*}
    &\pdv{S}{\alpha} = -1, \\
    &\pdv{S}{\beta} = -x_i, \\ 
    &\pdv{S}{\gamma} = -\cos(\omega x_i), \\ 
    &\pdv{S}{\delta} = -\sin(\omega x_i), \\ 
    &\pdv{S}{\omega} = +x_i\gamma \sin(\omega x_i) - x_i\delta\cos(\omega x_i).
\end{align*}
Without the last equation, it is a linear problem. As the gyrofrequency $\omega= q B/m$ only depends on the value of $B$ during its full orbit, the gyrofrequency can be estimated by taking the mean value of $B$ during the orbit. 

The expression of the function can then be rewritten as 
\[f(x,\vb*{\beta}) = \sum_{j=1}^4 \beta_j \phi_j(x) = \vb*{\beta} \cdot \vb*{\phi}\]
with 
\begin{align*}
&\phi_1(x) = 1 \\
&\phi_2(x) = x \\
&\phi_3(x) = \cos(\omega x) \\
&\phi_4(x) = \sin(\omega x) 
\end{align*}
Assuming $N$ observations, the observation matrix $X_{ij}:= \phi_j(x_i)$ is then 
\[X_{ij}= \begin{bmatrix} 1 & x_1 &\cos(\omega x_1)&\sin(\omega x_1)\\
\vdots &\vdots & \vdots &\vdots \\ 
1 & x_N & \cos(\omega x_N)&\sin(\omega x_N)
\end{bmatrix}.\]
The quantity $\vb{X}^T \vb{X} = X_{ji} X_{ik}$ can then be calculated as 
\[
\begin{bmatrix}
N & \sum x_i & \sum \cos(\omega x_i) & \sum \sin(\omega x_i)\\
\sum x_i & \sum x_i^2&\sum x_i \cos(\omega x_i) &  \sum x_i \sin(\omega x_i)\\
\sum \cos(\omega x_i) &\sum x_i \cos(\omega x_i) &\sum \cos^2(\omega x_i) &  \sum \cos(\omega x_i) \sin(\omega x_i)\\
\sum \sin(\omega x_i) & \sum x_i \sin(\omega x_i)& \sum \cos(\omega x_i) \sin(\omega x_i)&\sum \sin^2(\omega x_i) \\
\end{bmatrix}\]
The inverse of this can be calculated using Cramer's rule. The quantity $\vb{X}^T \vb{Y}$, where $\vb{Y}=y_i$, is 
\[\vb{X}^T \vb{Y} = X_{ji}y_i = \begin{bmatrix}
\sum y_i \\ 
\sum x_i y_i \\ 
\sum \cos(x_i) y_i \\ 
\sum \sin(x_i) y_i \\
\end{bmatrix}\]
The solution for the parameter vector $\vb*{\beta}$ which minimizes the sum of squares of residuals can now be obtained by \cite{davidson2004econometric} \[\vb*{\beta} = (\vb{X}^T \vb{X})^{-1}\vb{X}^T \vb{Y}\]
Then, the linear part of this gyromotion can be used to project the power exchange at the $(\mu, v_\parallel)$ location of the non-oscillating gyro-centre particle corresponding to the full-orbit particle (either at a few points if the $v_\parallel$ trend is steep, or only at the average, which proved to be sufficient in the cases considered in this work). 
\begin{figure*}
    \centering
    \includegraphics[width=\linewidth]{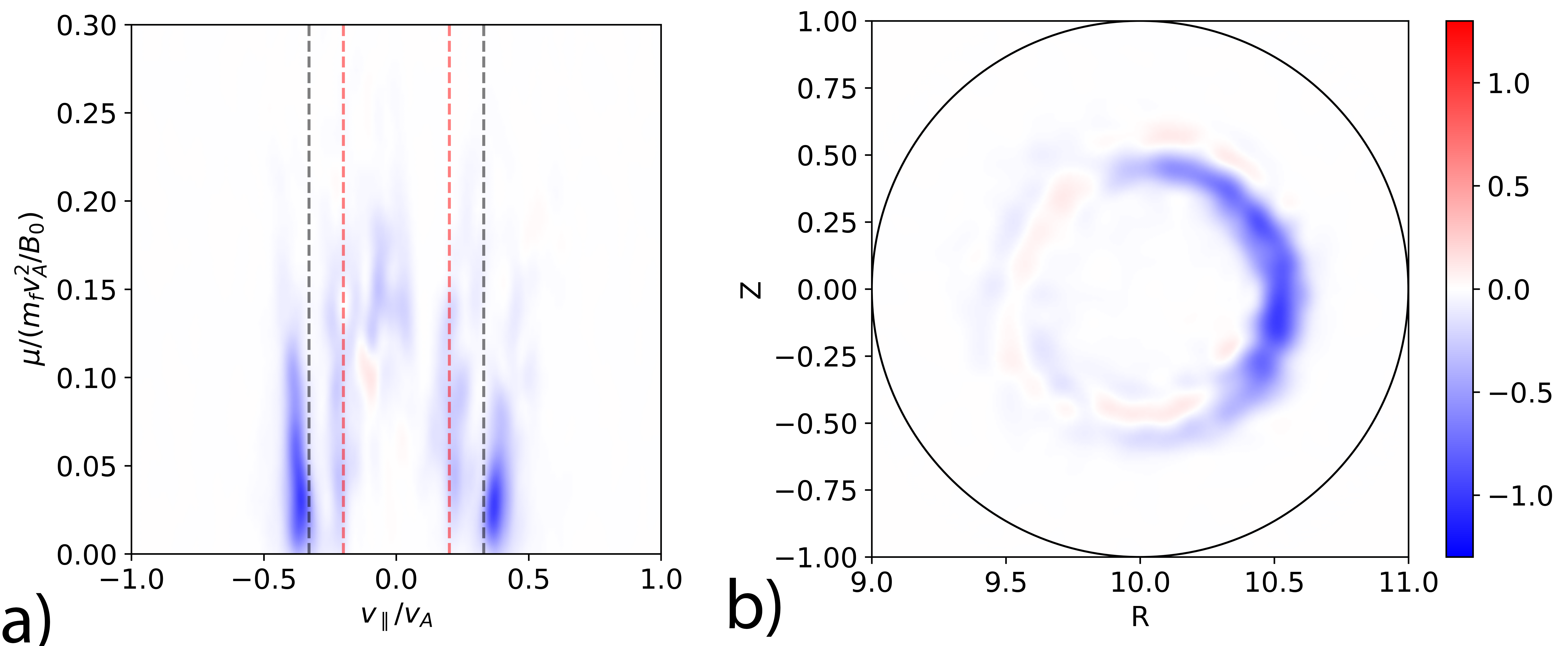}
    \caption{Energy gain or loss of the EPs in $(\mu, v_\parallel)$ space (a) and $(R,Z)$ space (b) in the ITPA-TAE benchmark. Here, the EPs have been fitted to constant $\mu$ and linear $v_\parallel$ variation, as outlined in the text, showing how this procedure allows for the visualisation of actual resonances without the dipoling effects.  In figure (a) the dashed lines indicate theoretical resonances, with black being $v_A/3$ and red $v_A/5$.}
    \label{fig:monopole}
\end{figure*}
The effect of this procedure is shown in figure \ref{fig:monopole}, where now only actual resonances are visible. \medbreak
A final remark is that it is very useful to sum (or average) the power exchange over many fluid timesteps, as a single fluid timestep only contains about five gyromotions, which is insufficient for the power exchange diagnostic.




\bibliography{aipsamp}

\end{document}